\documentstyle[amssymb,aps,epsf,rotate]{revtex}

\newcommand{\Label}{\label}
\newcommand{\Ref}{\ref}

\newcommand{\be}{\begin{equation}}
\newcommand{\ee}{\end{equation}}
\newcommand{\bna}{\begin{eqnarray}}
\newcommand{\ena}{\end{eqnarray}}

\renewcommand{\d}{{\rm d }} 
\newcommand{\half}{\frac{1}{2}}

\newcommand{\eps}{\epsilon} 

\newcommand{\CC}{{\mathbb C}} 

\newcommand{\NN}{{\mathbb N}} 
\newcommand{\RR}{{\mathbb R}} 
 
\newcommand{\ZZ}{{\mathbb Z}} 

\newcommand{\Ac}{{\cal A}}   
\newcommand{\Bc}{{\cal B}}   

\newcommand{\Dc}{{\cal D}}  %

\newcommand{\Lc}{{\cal L}} 
\newcommand{\Mc}{{\cal M}}

\newcommand{\Pc}{{\cal P}}

\newcommand{\Sc}{{\cal S}}

\begin{document}

\title{Negative and Nonlinear Response in an Exactly Solved  Dynamical Model of  
Particle Transport} 
\author{J. Groeneveld$^{1}$ and R. Klages$^{2}$} 
\address{$^1$Institute for Theoretical Physics, University of Utrecht,
Princetonplein 5,\\ 
3508 TA Utrecht, The Netherlands\\ 
e-mail: J.Groeneveld@phys.uu.nl\\
$^2$Max-Planck-Institut f\"ur Physik komplexer Systeme,         
N\"othnitzer Str. 38, \\
D-01187 Dresden, Germany\\
e-mail: rklages@mpipks-dresden.mpg.de}

\date{\today} 
\maketitle

\begin{abstract}
We consider a simple model of particle transport on the line $\RR$ defined by
a dynamical map $F$ satisfying $F(x+1) = 1 + F(x)$ for all $x \in \RR$ and
$F(x) = ax + b$ for $|x| < \half$.  Its two parameters $a$ (`slope') and $b$
(`bias') are respectively symmetric and antisymmetric under reflection $x \to
R(x) = -x$.  Restricting ourselves to the chaotic regime $|a| > 1$ and therein
mainly to the part $a>1$ we study, along the lines of previous investigations
[R.~Klages and J.R.~Dorfman, Phys.~Rev.~Lett. {\bf 74}, 387 (1995)] on the
restricted, symmetric ($b=0$) one-parameter version of the present model, the
parameter dependence of the transport properties, i.e.\ not only of the
`diffusion coefficient' $D(a,b)$, but this time also of the `current'
$J(a,b)$. A major  difference however is that this time an important tool for
such a study has been available, in the form of exact expressions for $J$ and
$D$ obtained recently by one of the authors. These expressions allow for a
quite efficient numerical implementation, which is important, because the
functions encountered typically have a fractal character.

The main results of our present preliminary survey of the parameter plane of
the model are presented in several plots of these functions $J(a,b)$ and
$D(a,b)$ and in an over-all `chart' displaying, in the parameter plane, in
principle all possibly relevant information on the system including, e.g.\,
the dynamical phase diagram as well as, by way of  illustration, values of
som topological invariants (kneading numbers) which, according to the formulas,
determine the singularity structure of $J(a,b)$ and $D(a,b)$.  What we regard 
as our most significant findings are:\\ 
 1) `Nonlinear Response': The parameter dependence of these transport
properties is, throughout the `ergodic' part of the parameter plane
(i.e.\ outside the infinitely many Arnol'd tongues) fractally nonlinear.\\ 
 2) `Negative Response': Inside certain regions with an apparently fractal
boundary the current $J$ and the bias $b$ have opposite signs.\\[2ex]
\end{abstract}

\section{Introduction}

 In many branches of science and even in mathematics, the study of highly
simplified models, alongside with a general theory on the particular subject,
is recognized as of great value. Such a model (or `toy' model) especially if
it can be analyzed in great detail, can serve as an illustration of the
general theory, can suggest further directions of development thereof and, if
the model exhibits unusual counter-intuitive behaviour, may even make a
revision of some tacitly made assumptions of the general theory necessary.  In
Dynamical Systems Theory, a rich source of such simple toy models are those
models the dynamical map of which is a `Lifted Circle Map' (`LCM').
 
 By `Lifted Circle Map' we will understand here 
just any real-valued function $F$
on $\RR$ satisfying for all $x\in \RR$ the relation

\be F(x+1) = 1 + F(x) . \Label{rel:ttr} \ee

 Maps of this kind, however under the restriction of having to satisfy
some further continuity condition, have been named, by Misiurewicz,
`old', as an acronym of `Lifted map of Degree One'.  However, we will
not use that name already because for the time being we will not
impose, unless explicitly mentioned, any further restriction other
than Eq.(\Ref{rel:ttr}).

 By a `LCM model' we will now understand a dynamical model, meant to be that of
a physical system, which is not only defined by a phase space $X$ and a
dynamical map $f$ of $X$ into itself, but also some further structure  
which is  invariant under the action of $f$; 
and possibly also a function $v$ on
$X$ such that $v(x)$ is the outcome of a measurement on the system if the
system's representative point is located at the point $x$ in phase space.

The theory of `LCM models' has a long history, dating back at least as far as
1895 when Poincar\'e (Cf.\ \cite{Ito}) defined the `rotation number' $\rho(F)$
for the subclass of the LCM's which are `orientation preserving
homeomorphisms', i.e.\ strictly increasing, continuous and with a continuous
inverse. Poincar\'e's definition of $\rho\equiv \rho(F,x_0)$ was by the
following limit (if it exists):

\be \rho(F,x_0) = \lim_{t \to \infty} x_t/t \ee

where the $x_t$ for $t \geq 1$ are defined in terms of $x_0$ by the dynamical
equation ($t \geq 0$)

\be x_{t+1} = f(x_t)  . \Label{rel:dyn} \ee

 A consequence of Poincar\'e's restriction was that  this limit then
always exists and that its value is independent of $x_0$, but it made
such maps also rather uninteresting from a physical point of view
because such models could not display diffusive behaviour. In
hindsight it may be surprising, but it was not before 1982 that
physicists \cite{GF1,GF2,GeNi,SFK} discovered that, without the
restriction of monotonicity, models of this class could also display
diffusive behaviour. Going now in a few big steps through the
subsequent history of the subject, we mention only some crucial
developments which led up to the present work and place the latter in
a certain context.

 At first, after this discovery, attention naturally focussed
on the determination of the diffusion coefficient of various such dynamical
models.  This was done both numerically and also exactly by mathematical means
\cite{GF1,GF2,GeNi,SFK,Cvi,Gasp98,Dor99}.  As it turned out, the exact answers 
were only obtained for piecewise linear maps, and for these the method which
was used was one of finding close by in parameter space a map having a Markov
partition and then performing the necessary algebra on the corresponding
Markov matrix.

It then turned out that such exact results always corresponded to isolated
points in parameter space, whereas at the same time, with the increasing
number of cases where such exact answers were derived, the parameter
dependence of $D$ started to look more and more complicated, and hence
interesting.  It was at this point that a determined effort was undertaken, by
J.R. Dorfman and one of the authors, to try to improve, in some way, the
existing techniques for solving this problem so that also cases with Markov
partitions of increasingly higher orders could be handled efficiently enough
so that the fine structure of the parameter dependence of $D$ could also be
determined.  As a model for applying their technique on, these authors then
chose a particularly simple model, which is the symmetric $b=0$ version of the
two-parameter model of the present paper. What they then discovered was, a.o.,
that this diffusion coefficient, $D(a,0)=c_2(a,0)$ in the notations of this
paper (Cf., e.g.\, Eq.\ (\Ref{expr:c2g}) or Eq.\ (\Ref{expr:c2s})), is a
continuous but fractal function of the parameter $a$ \cite{RKD,RKdiss,KD99}.

The present work is in a sense a continuation of the latter work, differences
however being that this time use is made of the set of exact expressions for
these transport properties such as $D$, and that the model has one more system
parameter, $b$, also called `the bias', which makes it possible to study now
also symmetry breaking phenomena, analogously to the theory of phase
transitions in Equilibrium Statistical Mechanics.

\subsection{Related results in the literature} \Label{ssect:RelLit}

1.  As it turned out, after the exact results on this natural one-parameter
extension of the original model of \cite{RKD,RKdiss,KD99} had been derived, it
was found that it contained, as another one-parameter specialization, also
another model which is extremely simple and a classic model, known from the
text books, carrying names such as the `Beta transform model' \cite{Flatto} or
the `Ren\'yi map model' \cite{Antoniou}.  These models are not obviously LCM
models but many of their properties relate to similar properties of its
extension, the present model.
 
2.  One of the results in Ref.\ \cite{Flatto} is particularly relevant for our
present discussion since it contains a result analogous to the fractal
parameter dependence of $D$ found in \cite{RKD,RKdiss,KD99}. It is the result
that also in that other one-parameter model a quantity was found to have such
behaviour of being continuous but obviously having a fractal nature. The
quantity in that case was the time of slowest decay.  The continuity result
stated in the present paper comprises and considerably generalizes both
results, those of \cite{RKD,RKdiss,KD99} and of \cite{Flatto}. The proof of
that generalization, as described in Subsection \Ref{ssect:ContP}, follows
closely that of the latter reference.
  
3.  Another related work which should be mentioned here is the work of Mori
\cite{Mori}, whose result is in one respect more general than the exact result
for the Fredholm determinant function $D(\lambda,u)$ of Section
\Ref{sect:Sol}, but in another respect a more special one, as will be
discussed further below in Section{\Ref{sect:PL}.

\subsection{Outline of the paper}\Label{ssect:OutP} 

An outline of the remaining parts of this paper is as follows: We will
first, in Section \Ref{sect:FoCo}, introduce successively some of the
notations and concepts to be used in the sequel. This starts with a)
cumulants, then b) the concept of the `long term behaviour' of a
stationary stochastic process, which is then c) extended to a process
generated by a dynamical variable on an `abstract dynamical
system'. This concept will then d) turn out to contain the information
one is most interested in in nonequilibrium theory, i.e.\ the
transport coefficients in the case of near-equilibrium states.

All of these considerations are intended to be only `formal' in the
sense that no conditions of validity are stated let alone that
mathematical proofs would be provided.  But they are also,
intentionally, extremely general.

 One reason for trying to be that general is that this may sometimes
lead to simplification in the presentation of a problem and then may
make a solution easier rather than more difficult to find.

 Along these lines, e) the formal considerations are continued until
the problem of calculating the properties of equilibrium and of
near-equilibrium transport is reduced to that of calculating a
particular (`weighted') Fredholm determinant.
 Subsequently, in section \Ref{sect:CGeom}, phase space is restricted
to be just the circle, or its `lift', the line $\RR$, in which case
the class of dynamical models arrived at is that of the so-called
`Lifted Circle Map models' or `LCM' models mentioned above.

In this case the programme can be carried through one step further, also
because of the special form taken by the weight function.

 It is shown that the complicating feature, that of the `weight'
associated with the Perron-Frobenius (from now on abbreviated as `PF')
operator can then be moved `out of the way' by transferring to a
representation of probabilities on the line $\RR$ where the
probability distributions satisfy a quasiperiodicity condition
(Cf. Eq.\ (\Ref{rel:quasip}) below).

In Section \Ref{sect:PL} a further specialization is made, to piecewise
linearity of the dynamical map $F$, in which case the programme has been
carried through to its end and an explicit expression is obtained for a
weighted Fredholm determinant which is relavant to the problem.  It contains
the information on the above mentioned properties of equilibrium and of
`near-equilibrium' transport, but also on time dependent properties such as
autocorrelation functions.  These results will be presented elsewhere
\cite{JG:tbp}.

Then finally, in Section \Ref{sect:PL1}, via a last specialization, our
special two-parameter `toy' model is reached on which, from then on, all our
attention will be focussed.
  
 In Section \Ref{sect:Sol} the centrally important `Consistency
Function'  will be constructed for this model, 
leading to the explicit expressions
for its  transport coefficients.

 Next, in Section \Ref{sect:Corrlrs} some of the  direct corollaries of
these latter results will  only be touched upon. These 
involve   two aspects of the solution: a) continuity properties of
the transport properties 
and b) questions of the occurrence of
Markov partition points in parameter space.
 
 The next Section \Ref{sect:Arnld} contains a summary of the formulas
specifying the boundaries of the Arnol'd tongues in the entire range
$a>0$. These results comprise results of Ref.\ \cite{Ding} for the non-chaotic
regime and our own ones for the chaotic regime.  The two collections of
formulas are written in a united notation which brings out some correlations
between the two kinds of results.  Both collections of results were
instrumental in preparing part of our final Figure \Ref{E7}, which displays the
most significant properties of the model.

In the subsequent Section \Ref{sect:Resp} we discuss the various types of
response encountered in our model system.

Then, in Section \Ref{sect:Figs}, we discuss in detail the seven figures
attached to the paper.

The paper then ends with a discussion of the results in Section
\Ref{sect:Disc} and with a summary of the results in Section
\Ref{sect:Summ}.

\section{Formal concepts } \Label{sect:FoCo} 
In this Section we will introduce some formal notions starting from well-known
ones such as that of a `cumulant', ending up with that of the weighted
Fredholm determinant associated with a dynamical variable on an `Abstract
Dynamical System', the determination of which is of the highest interest in
physical applications of Dynamical Systems Theory: Many properties of
dynamical systems, in equilibrium as well as in non-equilibrium states, could
be obtained if an efficient way of calculating that Fredholm determinant
function could be found.  Whereas it might seem far-fetched to expect a
solution of a problem that far-reaching, it should be noted that there already
exist several exact partial solutions of that centrally important general
problem, one of which is the fundamental formula of Gibbs for the equilibrium
state of a Hamiltonian system, the importance of which for Equilibrium
Statistical Mechanics need not be stressed.
  
\subsection{Cumulant rates and long term properties of a Stochastic
Process} \Label{ssect:CR-LTB}

We recall that the $n$-th order cumulant $\kappa_n(v)$ of a random
variable $v$ is defined by the formal relation

\be \sum_{n=1}^\infty \kappa_n(v) u^n / n! = \log <e^{uv}>
\Label{exp:kappa} \ee

A natural generalization of this concept is to apply it in a
particular manner to a stationary real-valued stochastic process

\be \vec{v} \equiv \{v_t|t \in \ZZ \} . \ee 

To this end we `integrate' this process to a new process $\vec{V}
\equiv \{ V_t|t \geq 0\}$ by defining, for $t \geq 0$, $V_t$ as the
sum

\be V_t = \sum_{s=0}^{t-1} v_s .\Label{def:V} \ee

By then defining $Q_t(u\vec{v})$ by

\bna Q_t(u\vec{v}) &=& <e^{u V_t}> \\
             {} &=& \exp({\sum_{n=1}^\infty \kappa_n(V_t) \, u^n/n!})
\ena

and taking the limits

 \be c_n(\vec{v}) = \lim_{t \to \infty} \frac{1}{t} \kappa_n(V_t) / n!
 \Label{def:cnlim} \ee

and
 \be c(u\vec{v}) = \lim_{t \to \infty} \frac{1}{t} \log Q_t(u\vec{v}) ,
 \Label{def:culim} \ee

and assuming that taking limits and series expansion commute here, we also
have
 
\be c(u\vec{v}) = \sum_{n=1}^\infty c_n(\vec{v}) u^n . \Label{exp:cu}
\ee

We will refer to $c_n(\vec{v})$ as the $n$-th order `scaled cumulant'
or `cumulant rate' and to $c(u \vec{v})$ as the `scaled cumulant
generating function' of the process $\vec{v}$.

 On this generalization or extension to stochastic processes we note
the following:
  
 1.  The terminology is consistent with the name `cumulant density'
used in Ref.\ \cite{Gasp98} for the related concept where the average
is taken over space instead over time as is the case here.

 2. The concept is a generalization
 or extension to stationary stochastic processes because, in the
special case of a process consisting of independent identically
distributed random variables, the two concepts coincide (apart from a
conventional factor $n!$ in the case of the $n$-th order
 cumulants.).

 3. A difference between the extension and the original concept is
that, whereas the set of cumulants of a random variable determines, in
a large of cases, the distribution of the random variable uniquely,
that is no longer true for a stochastic process.

 4. The present extension is useful because, in the case of a
stationary stochastic process, its scaled cumulant generating function
 will contain, in a precise manner, a type of information on the
process which one would call (as we will do) `the Long Term Behaviour'
(`LTB') of the process.
 
 5. The function $c(u\vec{v})$ is, by itself, well-known in many
fields and then may appear under many different names.

\subsection{Long term behaviour (LTB) in Abstract Dynamical Systems}
\Label{ssect:LTB3}

 Next we consider the case that the stationary process of the
preceding subsection is generated by a dynamical process, or rather by
observations on it in terms of a phase function $v$.
  
Accordingly, we will now suppose that we are given a set $X$ (`phase space'),
that $\Sc$ is a $\sigma$ algebra of subsets of $X$, that $\Mc$ is the linear
space over the complex numbers $\CC$ spanned by the measures on $(X,\Sc)$ and
that $f$ (`the dynamical map') is a $\Sc$-measurable map of $X$ into itself.

The motion through $X$ of a phase point representing the system is
assumed to proceed according to the dynamical law Eq.\
(\Ref{rel:dyn}).

This then defines our Abstract Dynamical System.

We also assume that $v$ (`phase function' or `dynamical variable') is
a real-valued $\Sc$-measurable function on the measurable space
$(X,\Sc)$.

Denoting now also by $\mu_t \in \Mc$ the measure determining the probability
distribution of the position $x_t$ of the representative phase point of our
dynamical model at a time $t$, the movement of the phase point as given by
Eq. (\Ref{rel:dyn}) causes the successive measures $\mu_t$ to be related by
the linear equation:
  
\be \mu_{t+1} = \Lc_f \mu_t  , \ee

where $\Lc_f$ is called the Perron-Frobenius (`PF') operator induced
by $f$.  This operator induces a linear transformation on the space of
measures on $\Mc$.

An important problem now is to determine, in the present case of our general
dynamical model, whether the above limits in Eqs. (\Ref{def:cnlim}) and
({\Ref{def:culim}) exist and if so, whether they are then independent of
$\mu_0$ (in which case they will be written respectively as $c_n(v)$ and
$c(uv)$), and then also whether the interchange of limits as expressed by Eq.
({\Ref{exp:cu}) holds true.

\subsection{The formal weighted Perron-Frobenius operator}
\Label{ssect:FWPFO}
 
 Continuing in the same formal manner we may write the function
$Q_t(u|\vec{v})$ of the previous subsection in the form

\be Q_t(u\vec{v}) = <\alpha(uv) | \Lc_f(uv)^t | \beta(uv)> \ee

for some nonvanishing linear functional $<\alpha(uv) |$ and -- function $ |
\beta(uv)>$ which are independent of $t$.

Herein $\Lc_f(uv)$ is a weighted Perron Frobenius operator which is related to
the original, `unweighted' one, $\Lc_f$, by

\be \Lc_f(uv) = \Lc_f e^{uv} \Label{rel:Lfu}  \ee

where  $v$ is the operator which, in its coordinate
representation on $X$, is equivalent to
 multiplication by the  phase  function $v$ of  the preceding subsection 
\Ref{ssect:LTB3}.

 Again, the above formulas can be made plausible as `formally valid' by noting
that there exist nontrivial examples for which they are valid.
  

 If we  now make  the further  assumption 
that, for $u$ real and $|u|$ sufficiently small, the above weighted PF
operator is irreducible and has an integral kernel in its  
`$X$-representation'
which is nonnegative, this  would allow application of  the PF
theorem,    leading to the  conclusion  that this operator has a unique
positive eigenvalue $\lambda_0(uv)$, to be referred to  
as 'the PF eigenvalue' (of this weighted PF-operator).
 
 For small enough $|u|$, this eigenvalue then would also be the
largest one in absolute magnitude, leading then to the conclusion that
$c(uv)$ would be expressible by the relation

\be c(uv) = \log \lambda_0(uv) ,\ee

whereas $\lambda_0(uv)$  would 
be a root of  an equation of the form

\be D(\lambda_0(uv),uv) =0 \Label{impl-eq:D}  \ee

with $D$ a  function  expressible in terms of a 
determinant, in the form:

\be D(\lambda,uv) = \det(1 - \frac{1}{\lambda} \Lc_f(uv))
. \Label{def:D} \ee
  
 This function $D(\lambda,uv)$ here would, in some sense, be `the
Fredholm determinant' of the operator $\Lc_f(uv)$.

 Furthermore, this root $\lambda_0(uv)$ would then be uniquely
determined, among all roots of the equation (\Ref{impl-eq:D}), by the
condition

 \be \lambda_0(uv) \to 1 \mbox{ \qquad for \qquad } u \to 0 . \ee

This then would finally have reducd the general problem, i.e.  that of
determining the equilibrium state and the near-equilibrium transport
properties of a general dynamical system, to that of the calculation, in an
efficient manner, of the above Fredholm determinant function $D(\lambda,uv)$
(we refer here also to a similar description in Ref.\ \cite{Dor99}, especially
Section 13.5, where also the corresponding references to the literature can be
found).

\subsection{On how to proceed from here, in the general case}
\Label{ssect:HowtoPr}

We now consider briefly, in passing, before we specialize to
one-dimensional phase space, what options there might exist for
proceeding in this general situation towards developing a feasible
generally applicable calculational method.

It seems that the following are at the moment the most promising ones:

1.  To expand, in Eq. ({\Ref{def:D}), the determinant function into inverse
powers of $\lambda$ and calculating, as far as possible, its coefficients
which then have the form of weighted sums over periodic orbits of the map $f$.
This is, in essence, the Periodic Orbit Expansion method \cite{Cvi}.  In
principle this method is generally applicable, but this statement should be
taken only in a `formal' way until also an efficient calculational scheme for
applying it in a general situation has been found.
  
2. Designing a systematic algorithmic approximation method for
arbitrary dynamical systems in terms of (approximate) Markov
partitions of phase space and then proceeding according to that
method.

 Although there seems to be nothing in the Markov partition method
which would put any restriction on the dimension of phase space, a
technique for applying it in a general situation appears not to be
available at the time.  Hence, the same remark as above seems to apply
here too: Practical application may crucially depend on the
construction of an efficient algorithm.
 
3.  For completeness' sake we note that the first order cumulant rate $c_1(v)$
is a linear functional whose knowledge is equivalent, as mentioned already in
the previous Subsection \Ref{ssect:CR-LTB}, to knowing the equilibrium state
of the sytem.  The remark now is that we only have to specialize to
Hamiltonian systems to arrive at a case in which the general problem already
has been solved exactly and explicitly long ago, in the form of Gibbs's
formula for the equilibrium state of a Hamiltonian system.
 
In order to make progress we will now impose, in the next section, the
drastic restriction on the phase space of being one-dimensional.

Because of our motivation to study transport, it is important for us
to have a phase space which is not simply connected, but neither has
singularities; hence, there is no other
 possibility than to choose for this the circle.


 
\section{Circle geometry} \Label{sect:CGeom}

As a preparation, we will recall some simple circle geometry.

\subsection{Preliminaries} \Label{ssect:Prel}

The definition of a LCM has already been given in the introduction. It is that
of a real valued function on $\RR$ satisfying Eq.\ (\Ref{rel:ttr}).

We introduce now also the `unit translation map' $T$ by

\be \mbox{for all } x \in \RR \; : \; T(x) = 1 + x . \Label{def:T} \ee

Then the above defining relation (\Ref{rel:ttr}) can equivalently be written
as the `commutation relation'

\be FT = TF \Label{rel:FTTF} . \ee

The set of real numbers $\RR$ regarded as an additive group is the universal
covering group of the circle considered as the factor group $\RR/\ZZ$.

Both $\RR/\ZZ$ and $\RR$ may also be regarded topologically, in which case the
latter is the universal covering space of the former.

\subsection{Splitting a real number} \Label{ssect:spl1}
 
Now we introduce some notation which we need for clarifying the relation
between a circle (with circumference one) and the real line.

For the convenience of constructing a corresponding coordinate system, we
choose a fixed half-open half-closed unit sub-interval of $\RR$, $I_0$.

In terms of this, an arbitrary real number $x \in \RR$ can be decomposed
uniquely into a pair of numbers 

\be x \rightarrow (y,n) \ee 

by the relation

\be x = y + n \Label{split:x} \ee 

and the conditions 

\be y \in I_0 \mbox{\qquad and \qquad} n \in \ZZ . \Label{cond:x} \ee

We will also write this as 

\be y = \pi(x) \mbox{\quad ,\quad} n = \sigma(x) , \ee 

wherein, obviously, $\pi$, and therefore also $\sigma$,
 is a projection map.

\subsection{Splitting a lifted circle map} \Label{ssect:spl2}
We define also the translated unit intervals $I_n$ by 

\be I_n = T^n I_0 . \ee

Then, analogously to the splitting of a real number, also an arbitrary LCM $F$
may now be split in a unique manner into a pair of functions according to

\be F \rightarrow (f,m) \ee
 
wherein these two functions are defined by the relations

\be F = f + m \Label{decomp:F} \ee

and the conditions

\be \mbox{if \quad} x \in I_n \mbox{\qquad then \qquad} f(x) \in I_n \ee

and

\be m(x) \in \ZZ .\ee 

It then follows readily that

\be mT = m \Label{rel:decomp} , \ee

i.e. $m$ is a $\ZZ$-valued periodic function with period $1$, and

\be fT = Tf , \ee

i.e. $f$ is a LCM of a special class, one which maps every $I_n$ onto
itself, and that of course in an identical manner.

One can now also show that the map $F \to f$ is a projection, that the
above decomposition of an arbitrary LCM $F$ into a pair $(f,m)$ is
invertible and that, for the inverse map (`map of maps') from the pair
$(f,m)$ the two components can be chosen independently of one another.

Because $f$ can be obtained from $F$ by a projection, $F$ is called,
as is usual, a `lift' of $f$.  The function $m$ determines, in this
context, to which function $F$ the function $f$ is lifted. It may
therefore appropriately be called the `lift function'.

\section{The abstract lifted circle map model} \Label{sect:ALCMM}

 We now discuss the class of dynamical models (termed `LCM models')
whose dynamical map is a LCM.
  There are two ways of considering what its phase space $X$ in this case is. 
It is either: 
 I.) the real line $\RR$ or II.) the circle $\RR/\ZZ$. 
 
 Such a model is specified by: 

 1.) the collection  of  subsets  of $\RR$ which are to be considered 
 'measurable'. This amounts to the choice of 
 a $\sigma$ algebra $\Sc$ of such  subsets of $\RR$.
This algebra should be chosen to be  $T$ invariant. 

 2.) a lifted circle map $F$, also   leaving $\Sc$ invariant. 
The   circle map $f$ `associated with $F$' via the choice of 
a fundamental interval $I_0$, as in Section \Ref{sect:CGeom}, 
will then also leave $\Sc$ invariant.

 Then to each of these maps $T$, $F$ and $f$ there will exist a 
corresponding  PF operator leaving  invariant 
the space $\Mc$ spanned by the  measures on $\Sc$; and 

 3.) an  `initial measure' $\mu_0$ chosen from $\Mc$.
Without lack of generality, this choice  can be made  so as 
to   vanish outside of $I_0$.
 
\subsection{Simplification by Gauge Symmetry: First step}
\Label{sect:GaugeS}
 
 We now again take up the thread from Subsection \Ref{ssect:FWPFO} where we
left it.

Due to the fact that for the calculation of  transport properties 
the function $v$ is no longer arbitrary 
but has the special form  given by 
 
\be v(x) = F(x) - x \ee

it is  now  possible  to rewrite the weighted PF operator of
\Ref{ssect:FWPFO} starting from relation (\Ref{rel:Lfu}) in successively
simpler forms as follows:

\bna 
\Lc_f e^{uv} &=& \Lc_f e^{u(F(y) - y)} \\
&=& \Lc_f e^{u(f(y) + m(y) - y)}\\ 
&=& e^{uy} \Lc_f e^{u( m(y) - y)} \\
&\sim&  \Lc_f e^{u m } \\
&=& \Lc_F^{(u)} 
\ena  
 where, in the next-to last equation, the similarity symbol denotes similarity
of operators, and where in the last line, the superscript $(u)$ refers to the
fact that the PF operator induced by $F$ is in that case acting on a space of
functions satisfying the quasiperiodicity condition 
(\Ref{rel:quasip}) below.

 In what we have termed the first way, (I),  of considering the system with the 
line $\RR$ as its phase space, 
the above sequence of identities 
can also be expressed by saying   that the `weight' on the
original PF operator has been transformed away, at the `expense' 
that the PF operator $\Lc_F$ (now written as $\Lc_F^{(u)}$ ) 
acts on a space of functions $\psi$, say, which
satisfy the `quasi-periodicity condition as in

\be \psi(x+1) = e^{-u} \psi(x). \Label{rel:quasip} \ee
 
 That such a transformation  
is possible can be seen to be due to the fact that 
the problem is, by its nature, essentially  a gauge
theory. 

 In the second  way, (II),  
of describing  the system, with as  phase space now 
the interval $I_0$ which is equivalent to a circle with 
 one special point on it, singled out, 
the above can be interpreted as
follows: The PF appropriate operator now moves probability along as
prescribed by the dynamic equation (\Ref{rel:dyn}) whereby  conserving
this probability, i.e. `locally'; but when that 
special point has to be  passed,  probabilities are
rescaled by a fixed factor $e^u$  or its inverse, depending on the direction of
passage. This can be expressed by saying that probability is conserved
`locally' but not `globally'; which are ways of expression quite 
familiar in Gauge Theory.
 This  representation on the interval then is said to be one 
`with twisted' boundary conditions.
 
\section{Specialization to the class of  general piecewise linear LCM's: final 
steps taken}
\Label{sect:PL}

We now further specialize to the class of LCM models whose dynamical map $F$
is piecewise linear.

For this class the calculation of the `twisted' Fredholm determinant function
of the PF operator $\Lc_f(u)$ has been carried through to the end, meaning
that the answer has been written in a form which is such that its computation
can be performed algorithmically with sufficient efficiency.  In this paper we
will now describe only an outline of what this final step consists of.
 
The results of this will then be described in the next Section \Ref{sect:Sol},
after further specialization to the simpler case that the number of `laps of
linearity' per period, $L$, is minimal, i.e.\ equal to one.  A proper
derivation in the case of general $L$ will be given elsewhere \cite{JG:tbp}.

This `final stage' of our derivation will now be described in words and will
take five steps:

1. The first step is to reinterpret the probability densities on the line
$\RR$ or the circle $\RR/\ZZ$ as electric field strength distributions in a
problem of one-dimensional electrostatics.

2. Then, in a second step, the calculation in terms of the electric field
strength distribution is replaced by one in terms of the derivatives of these
fields, which are then of course to be interpreted as `charge densities'.
This requires the original PF operator governing the movement of probability
densities of the original representation to be transformed accordingly.

An important circumstance here is that in the above translation from fields to
charges, which is simply that of differentiation, no information is lost,
provided that $e^u \neq 1$, because in the inverse process, which is that of
integration, the integration constant is fixed by the quasi-periodicity
condition Eq.\ (\Ref{rel:quasip}) on the result.

This charge density representation has the added advantage that the
correspondingly transformed PF operator tends to decrease the charge
densities, on average, at least in case the system is ergodic and mixing, by a
factor which is conjectured to be equal to the largest Lyapunov
exponent. Already intuitively one can therefore expect this process to
converge, as long as the system is ergodic and mixing. To this it  may be
added that the explicit form of the solution, which  is found this way, will
also specify explicitly the set of conditions under which the method will be
applicable.

3. The next step is to write down the possible form an eigenfunction of the PF
operator can have.  In the case of an $F$ having a finite number, $L$, of laps
of linearity per period, this requires $L$ as yet undetermined constants; but
then, in the end, one must impose, for each interval of linearity, an
independent condition requiring the total charge within the interval to
vanish.  Such charge neutrality can always be achieved by introducing a charge
double layer of appropriate strength at each point between two laps.

This leads to a  system of $L$ linear equations in as many unknowns, 
 the $L^2$
coefficients of which are recursively calculable  functions 
of $\lambda$ and $u$.

4. The condition of solvability of this system then leads to a single
condition on a determinant function (here termed `Consistency Function')  
for it to  vanish.

5. This determinant then can be seen to contain as one of its roots the
desired function $\lambda_0(u)$.\\
{\bf Remarks:}

1.  A proper, more explicit account of this derivation will be given elsewhere
\cite{JG:tbp}.

2. As was already mentioned in the Introduction, the result described here in
general terms generalizes Mori's result~\cite{Mori}, which is a formula for
the Fredholm determinant in the case of the general piecewise linear interval
map.

The generalization is one from interval maps to lifted circle maps, both in
the piecewise linear case. It is this more general formulation which allows
for the possibility of transport phenomena to occur.

In the next Section \Ref{sect:Sol} the solution in the case $L=1$ will be
treated explicitly; in which case the Consistency Function mentioned is the
function $C$ of Eq.\ (\Ref{def:C}).
 
\section{Specialization to the case $L=1$: a two-parameter model}
\Label{sect:PL1}

We specify such a map by two parameters $a$ and $b$ so that, accordingly, $F$
will now be specified completely by

\be F(x) = ax + b \Label{def:F-ab}, \ee
 
and specifying finally also the value taken on by $F$ at $x=\half$. The
latter number is chosen arbitrarily but will not enter into any of our
considerations below since these will be confined to `physical' quantities
such as the $c_n$'s, and the latter will not depend on that choice.

Hence, $a$ and $b$ are effectively the only parameters which specify the
system. They are termed the `system parameters'.

Our interest will primarily be the chaotic region characterized by
 
\be |a| > 1 .  \Label{ineq:a} \ee

As for notation: In case we want to consider different members of our
two-parameter set of models and of maps, we may denote the dynamical map $F$
specified above as $F_{a,b}$.

\subsection{`$\lowercase{b}$-symmetries'} \Label{ssect:Sym}
 
The present two-parameter problem obeys certain simple symmetries, and it is
advantageous to explicitly consider the corresponding symmetry group.

In addition to the map $T$ defined in Eq.\ (\Ref{def:T}) we now also introduce
$R$, the reflection map, by

\be R(x) = - x . \Label{def:R} \ee
  
These two maps generate a group as follows: A said, its generators are $T$ and
$R$, and these satisfy the relations

\be R^2 = \mbox{id} \ee

and 

\be TRT = R . \ee
 
The elements of this group belong to two separate classes: respectively $C_0=
\{T^n | n \in \ZZ\} $ and $C_1 = \{T^n R| n \in \ZZ \}$, with $C_0$ a normal
subgroup of the group.
 
This symmetry group occurs here as a transformation group of $\RR$ and it will
be of interest, as we shall see in the following, to determine the fixed
points.  We find that, with the exception of the identity no other element of
$C_0$ has a fixed point, whereas every element of $C_1$ has one: The element
$T^m R$ leaves the point $x = m/2$ invariant.  Considering now this collection
of these fixed points we see that, with respect to this group, there are two
classes, the integer and the half-integer numbers.
 
We can also classify all elements of $\RR$, which in this context has to be
regarded as the $b$-axis, with respect to this group, and we obtain the result
that the closed interval $0 \leq b \leq \half$ is a fundamental subset of the
$b$-axis with respect to this symmetry group.  Hence, the two boundary points
of this closed interval play a special r\^ole here.  As we shall see, it is
here that the model displays its most irregular behaviour.

We now list the effects which these symmetries have on the the two-parameter
set of maps $F_{a,b}$. We consider only the effect of these two generators $T$
and $R$:

1.  Translation symmetry:

\be \mbox{(T1)\qquad} F_{a,b} T = F_{a,b+1}, \Label{rel:FabT} \ee

cf. Eqs.(\Ref{rel:ttr}) and (\Ref{rel:FTTF}), and the

2.  Reflection symmetry:

\be \mbox{(R1) \qquad} F_{a,-b} = R F_{a,b} R .\Label{rel:FabR} \ee

It follows from this in a straightforward manner that, for all $n
\geq 1$,

\be \mbox{(T2) \qquad}c_n(a,b+1)=c_n(a,b)+\delta_{n,1} \Label{relT2}
\ee

and

\be \mbox{(R2) \qquad} c_n(a,-b)=(-1)^nc_n(a,b). \Label{rel:2} \ee

These symmetry relations imply that for the study of the model throughout its
entire parameter plane it is sufficient to restrict oneself to $b$-values 
within
the closed interval given by 

\be 0 \leq b \leq \half.\ee
 
This specifies what we call the `fundamental strip' in the parameter plane.

Parts of this `fundamental strip' are used in our Figs. \Ref{E1}, \Ref{E5},
\Ref{E6} and \Ref{E7}.

\section{Construction of the Consistency Function} {\Label{sect:Sol}}

In this section we construct the Consistency Function for our two-parameter
model.  As stated already, this function contains the Fredholm determinant
function of the problem and is of central importance in the solution: It
determines all of the `near-equilibrium' transport properties $c_n$ with in
particular $J$ and $D$, but also the spectrum from which time-dependent
quantities can be obtained.

\subsection{The kneading- and the $y$-sequences} \Label{ssect:Kndng}
  
Up to now we had considered one fixed `fundamental interval' denoted by $I_0$,
but now, for the explicit construction of the solution, we will need two such
`fundamental' intervals. If in the sequel reference will be made to $I_0$, or
indirectly to it by referring to `the associated circle map $f$' which is
defined on the basis of $I_0$, we will assume that $I_0$ is just either one of
the half-open unit intervals $I_0^\eps$ defined by

\be I_0^+ = (-\half,\half] \quad \mbox{and} \quad I_0^- =
[-\half,\half) \Label{def:I0eps} \ee
 
This puts us now in a position to also define recursively, for a given
parameter pair $(a,b)$ with $|a| > 1$ and each value $+$ or $-$ of $\eps$
separately, a pair of sequences of numbers, the $\vec{y}^\eps$ sequence and
the $\vec{n}^\eps$ sequence, 

\be \vec{y}{\,}^\eps = \{y_r^\eps | r \geq 0 \} , \Label{def:yvec} \ee 

and

\be \vec{n}{\,}^\eps = \{n_r^\eps | r \geq 1 \} \Label{def:nvec} \ee 

consisting of real numbers and of integers, respectively, as follows:

We start with

\be y_0^{\eps}=\frac{\eps}{2} \Label{def:y0eps} \ee

and then continue recursively, for all $r \geq 1$, by means of the relations

\be n_r^{\eps}+y_{r}^{\eps}=ay_{r-1}^{\eps}+b \Label{rel:nyr} \ee

and  the conditions

\be n_r^{\eps}\in \ZZ \qquad \mbox{ and } \qquad y_{r}^\eps \in
 I_0^{\eps \eta^{r}} , \Label{cond:nyreps} \ee

where the sign $\eta$ is defined by

\be \eta = \mbox{ sign}(a) \equiv a / |a| . \Label{def:eta} \ee

We note that the above implies that, for all $r \geq 0$, $\eps \eta^r y_r^\eps
\in I_0^+$.
 
The integers $n_r^\eps$ are analogous to the kneading numbers known
from the theory of maps of an interval onto itself
\cite{Milnor,Collet}.  In fact, they are topological invariants of the
map $F$ in the sense that, if we would perform an arbitrary
topological, i.e.\ continuous and continuously invertible
transformation on the circle, these numbers would not change. This
will play a r\^ole in the subsequent discussion of our results.
 
Next, for $\eps=\pm$ and all $r \geq 0 $, the ``$N$-numbers'' are
introduced by

\be N_r^{\eps}= -\frac{\eps}{2} + \sum_{1 \leq s \leq r}\! \! {}^{'}\,
n_s^\eps . \Label{def:N} \ee

\subsection{The Consistency Function $C(\lambda,u)$}
\Label{ssect:ConsF}
 
The `Consistency Function' $C(\lambda,u)$ which plays the key r\^ole in the
present context is now introduced by
 
\be C(\lambda,u)=\sum_{\eps=\pm}\eps\sum_{r=0}^{\infty}(a
\lambda)^{-r}e^{uN_r^{\eps}} \Label{def:C} .\ee

For future reference we note that, from Eqs.\ (\Ref{def:I0eps}),
(\Ref{def:y0eps}), (\Ref{rel:nyr}) and (\Ref{cond:nyreps}), the following
upper bound can be derived:
 
\be |n_r^\eps| \leq (|a| + 1)/2 + |b| \equiv B \Label{ineq:n}  \ee
   
which also introduces the constant $B$, implying that

\be |N_r^\eps | \leq \half + r B . \Label{ineq:N} \ee

It follows that series (\Ref{def:C}) converges for all $(\lambda,u)$
satisfying

\be B |\Re( u) | \: < \: \log  |a  \lambda|, \Label{ineq:lambd-u} \ee
 
throughout which region the function $C$ is therefore holomorphic.  Because of
the assumption $|a| > 1$, this region contains the point $(1,0)$ in its
interior.

Also, that same region, now considered as a $u$-region for a given $\lambda$,
contains, for all sufficiently large $|\lambda|$, a complete strip of the $u$
plane parallel to the imaginary $u$-axis; and because, as we can see from the
definition Eq.\ (\Ref{def:N}), $N_r^\eps \in \ZZ + \half$, $C$ is
antiperiodic, i.e.\ odd under the substitution $u \to u + 2 \pi i$, and it
vanishes at $u=0$. We also note that

\be C(\infty,u) = - 2 \sinh(\frac{u}{2}) \Label{lim:C}\ee

which implies that also the $D$-function defined by

\be D(\lambda,u) = C(\lambda,u) /C(\infty,u) \Label{def:D1} \ee

is holomorphic in the pair $(\lambda,u)$ throughout the same
$(\lambda,u)$-region and, in contrast to the $C$-function, periodic in the
above sense.
 
It seems significant to remark, concerning these formulas, that this
proportionality factor between these $C$ and $D$ functions vanishes whenever
$e^u = 1$.  Hence, in a way, this `weight' on the PF operator has been
instrumental in deriving also many  non-equilibrium properties of the system
when there is no `weight'.

The function $D(\lambda,u)$ can be regarded as being, in some sense, the
Fredholm determinant of the operator $\Lc_f(u)$ because \\ 
1) its limit is $1$ as $\lambda \to \infty$ and \\
2) for fixed $u$, its zero set coincides with that of $C$ which is, as can be
seen, the set of eigenvalues $\lambda$  of $\Lc_f(u)$ 
satisfying (\Ref{ineq:lambd-u}).
  
We summarize the preceding considerations with the remark that these crucial
functions $C$ or $D$ encountered here are not obtained by the expansion of a
determinant and subsequent calculation of traces of powers of an operator as a
sum over periodic orbits of a map, as in the case of the Periodic Orbit method
\cite{Cvi}, which are in a sense `direct' methods for obtaining the desired
answer, and which is also the way the zeta function is usually defined, but in
a roundabout, indirect manner, which however as yet is only applicable in the
case that $d=1$ and $F$ is piecewise linear, but then turns out to be
extremely effective.
 
Before we formulate the central result concerning our present model we
introduce the following notation: We denote by $\Pc$ the class of probability
measures $\mu$ on the line $\RR$ which have a density $\rho$ given by $\d
\mu(x) = \rho(x) \d x$, which  is a function of
finite total variation, i.e.\ $\rho$  
can be expressed as the difference between two
nondecreasing functions.  
 Then we have the\\
\noindent {\bf Theorem}~\cite{JG:tbp}\\
In the two-parameter model discussed here, with a dynamical map $F$ which
satisfies the relations Eq.\ (\Ref{rel:ttr}) and Eq.\ (\Ref{def:F-ab}) with
$|a| > 1$, and where the probability measure $\mu_0$ of the initial position
$x_0$ belongs to the class $\Pc$, one has:
\begin{enumerate} 
\item 
The limit 
\be c_n(\mu_0) = \frac{1}{n!} \lim_{t \to \infty} \frac{1}{t} <<(x_{t})^n>>_0 
\Label{def:cn} \ee 
exists for every $n\in\NN$ and is independent of $\mu_0$; it will
be denoted by $c_n$.
\item 
If $u$ is a complex number satisfying
 \be | \Re(u) | < (\log |a|) /B \Label{ineq:Reu} \ee
the limit 
 \be c(u|\mu_0) = \lim_{t \to \infty} \frac{1}{t} \log <e^{u x_{t}}>_0  \ee
exists and is independent of $\mu_0$; it will be denoted by $c(u)$.
\item
The function $c(u)$ is analytic in $u$ in the region
 \be |u| < (\log |a|)/B \Label{ineq:u} \ee
and has as its series expansion 
 \be c(u) = \sum_{n=1}^\infty c_n u^n \Label{exp:c} \ee
the coefficients in which are the same $c_n$'s as have occurred in Eq. 
(\Ref{def:cn}).
\item 
The analytic function $\lambda_0(u) = e^{c(u)}$ is a root of
 the implicit
equation
 \be C(\lambda_0(u),u) = 0 \Label{eq:impl} \ee
singled out among all roots 
by  the additional condition
 \be \lambda_0(u) \to 1 \qquad {\mbox as } \qquad u \to 0 \Label{as:l-u} . \ee

 \end{enumerate}

\subsection{$(c,u)$ expansion} \Label{ssect:cuExp}

Because of the above, the right hand side of Eq.\ (\Ref{def:C}) can be expanded
in a converging double series in $c\equiv \log \lambda$ and $u$ near $(0,0)$
according to

\be C(e^c,u)= \sum_{k,l=0}^\infty u^k N_{k,l} (-c)^l, \Label{exp:C} \ee

the coefficients of which are defined, for all $k$ and $l$, by the 
expansions

\be N_{k,l} = \sum_\eps \eps N_{k,l}^\eps ; \qquad   N_{k,l}^\eps = 
\frac{1}{k!l!}  \sum_{r=0}^\infty a^{-r}
(N_r^\eps)^k r^l \Label{def:Nkleps}. \ee

The above infinite series 
is guaranteed to converge because of  the assumption $|a| >
1$; the   left hand sides of these equations are referred to as  `$N$-moments'.

\subsection{Solving for the $c_n$'s} \Label{ssect:Resltn}

The implicit equation (\Ref{eq:impl}) for $c(u)$ can now be resolved
by standard mathematical means into a set of explicit ones, one for
each expansion coefficient $c_n$.
  To show that such a resolution is possible use is made of the
relations $N_{0,l}=0$ for all $l \geq 0$ which follow trivially from
the definition, of $N_{1,0}=0 $ which can be derived from the
recurrence relations and the definiton of
$N_{1,0}$, and of the circumstance that the quantity $N_{1,1}$ never
vanishs, which is a consequence of the important inequality

\be \qquad N_{1,1} > 0  \Label{ineq:N11} , \ee
 
which can be derived by application of the Perron-Frobenius Theorem (Cf.,
e.g., \cite{Gantmacher}) to the invariant eigenfunction of the
Perron-Frobenius operator $\Lc_f(0)$~\cite{JG:tbp}.\footnote{By a more refined
analysis also a positive lower bound for $N_{1,1}$ can be derived, which can
come in handy because of the divisions by this quantity which are required.}
 
The above inequality Eq.\ (\Ref{ineq:N11}) implies in particular that division
by $N_{1,1}$ is always possible, which then leads to the following equations
for the transport properties $c_1=J$ and $c_2=D$, which hold true for all
$(a,b)$ throughout the chaotic region $|a|>1$, and which, in particular, are
also independent of whether or not for the given pair $(a,b)$ a Markov
partition exists:
 
\bna J \equiv c_1 &=& N_{2,0}/N_{1,1} \Label{expr:c1g} \\ D \equiv c_2
&=& (N_{3,0} - N_{2,1} c_1 + N_{1,2} c_1^2)/N_{1,1}.  \Label{expr:c2g}
\ena

The expressions for the higher cumulant rates, explicit as polynomials in the
$N_{j,k}$ and $1/N_{1,1}$, become successively more complicated and are
therefore more conveniently formulated in a recursive form \cite{JG:tbp}.  In
the symmetric case $b=0$ however, the odd order $c_n$'s vanish and the
expressions for $c_2$ and $c_4$ read:

\bna c_2 &=& N_{3,0}/N_{1,1} \Label{expr:c2s} \\ c_4 &=& (N_{5,0} -
N_{3,1} c_2 + N_{1,2} c_2^2)/N_{1,1}.  \Label{expr:c4s} \ena

One may notice that these latter two expressions for $c_2$ and $c_4$ in the
symmetric case are  analoguous to the preceding ones 
for $c_1$ and $c_2$ in the general case. This analogy can be seen to persist
to general order~\cite{JG:tbp}.

\section{Corollaries} \Label{sect:Corrlrs}

We discuss here some of the consequences which can be derived already rather
effortlessly from the defining relations for the kneading sequences in
Subsection \Ref{ssect:Kndng} and the further relations of Section
\Ref{sect:Sol}.

The consequences we discuss here are about existence and density of Markov
partition points and about continuity or discontinuity of the various
functions encountered here.  These consequences are the following:

\subsection{Markov partition points} \Label{ssect:MPpts}

  There exist two collections, say $\Ac^+$ and $\Ac^-$, of algebraic curves,
both dense throughout the chaotic region $|a|>1$, such that any two 
curves of the same collection do not  intersect.

 For given $\eps$, the points on  a   curve of $\Ac^\eps$ are  
characterized by having the same $\eps$-kneading sequence 
$\vec{n}^\eps$. The latter  may therefore be termed `the kneading sequence of 
the respective curve'. 

 Each  curve of the  collection $\Ac^\eps$  has a  kneading sequence  
which is `eventually' periodic, i.e.\ periodic after a finite number 
of  `steps'.

 If two   curves of the two 
different collections intersect they always do so transversally. This
is a direct consequence of the fundamental inequality Eq.\
(\Ref{ineq:N11}).

{\bf Note.}  These arguments do not imply that, as one would expect,
any two curves of these two collections would intersect at all. To show that,
a more refined analysis would be required.

However, what does follow here is that the point set of intersections $\Ac=
\Ac^+ \cap \Ac^-$ coincides with the set of points for which the circle map $f_{a,b}$,
which is associated to the map $F_{a,b}$ on the line, has a (finite) Markov
partition.

\subsection{(Dis-)continuity properties} \Label{ssect:ContP}

 a.) We first discuss the parameter dependence of the `$N$-moments' in
terms of which the $c_n$'s are expressed. One readily derives the
following:

 There exist two collections, say $\Bc^+$ and $\Bc^-$, of algebraic
curves, which are subcollections of the respective collections $\Ac^+$
and $\Ac^-$, each of which is also, just as the collections
$\Ac^\eps$, dense throughout the chaotic region $|a|>1$.
 A collection $\Bc^\eps$ is obtained from $\Ac^\eps$ by restriction to
curves whose kneading sequences $\vec{n}^\eps$ are required to be
periodic.

The $N$-moments $N_{k,l}(a,b)$ (Cf. (\Ref{def:Nkleps})), regarded as
functions of $(a,b)$, are discontinuous when crossing any one of these curves,
but continuous everywhere else, i.e.\ everywhere outside the set $\Bc = \Bc^+
\cup \Bc^-$.
 
b.)   Considering  now   the   $c_n$'s:  
According to the relations of Section
\Ref{sect:Sol} they are expressible as rational functions of the
$N$-moments and hence could  have been expected 
to be discontinuous in the same
way as the $N$-moments -- unless, of course, a `miraculous' cancellation would
occur.  A cancellation, miraculous or not, does indeed occur because these
$c_n$'s themselves can be proven to be continuous throughout their domain of
definition, which is the chaotic region $|a| > 1$.
 
 The proof runs analoguously to one of the continuity property in Ref.\
\cite{Flatto}. A crucial ingredient in the proof here is that the only effect
upon the Consistency Function in Eq.\ (\Ref{def:C}) of crossing any one of
these `$\Bc$-curves' is a multiplication by a nonvanishing over-all factor.
Hence, this crossing does not affect any root of the equation (\Ref{eq:impl}),
nor the value of $\lambda_0(u)$ nor that of the analytic function element
$c(u)$ nor that of any other root of this equation within the domain of
definition $|a \lambda| > 1$ of the $C$-function. This same is expected to
apply to any time dependent property of the system.

c.) The continuity result on $J(a,b) \equiv c_1(a,b)$ has an important impact
on the way figures such as Figs.\ \Ref{E5} and \Ref{E6} must be produced.  It
means that the point sets to be displayed where $J$ or $\half - J$ should have
a particular sign must consist of 
open regions and therefore cannot be too `wild'.

\section{Phase locking regions (Arnol'd tongues) }\Label{sect:Arnld}
 
The phenomenon of phase locking in lifted circle maps is basically
well understood. In the case of the present model with $a>0$, 
the regions where this takes place  have
been determined exactly: In  the non-chaotic regime $0 \leq a \leq 1$
in Ref. \cite{Ding} and  in
the chaotic regime $1 \leq a$ they  are given here.

 A  region in parameter space where the system exhibits    
phase locking   can be interpreted as the dynamical analogue of a 
thermodynamic `phase';  `phase locking' then can be seen as 
 a case  of  `spontaneous symmetry breaking', the symmetry 
being then  that of `time translation'.

  In our present  
model we are not aware whether  there would exist 
 any other `dynamical phases' than `phase locking' or `no phase locking'.

Therefore, the subdivision of the parameter plane into these
Arnol'd tongues and their complement,  which latter could   be 
called  `the ergodic phase', deserves  the name 
`Dynamical Phase Diagram' of the model.

 This forms an important  part of 
our display of the positive-$a$ part of the
parameter plane in Figure \Ref{E7}.

\subsection{The boundaries of the Arnol'd tongues in the two $a$-
regimes} \Label{ssect:ArnldB}

In formulating these results we adopt the following further notation:

In considering any particular Arnol'd tongue inside of which the fixed value
of $J$ is equal to the rational number $q:p$, we will adopt the convention
that $p$ and $q$ are relatively  prime integers with $p$ positive. This we call
the `standard convention'.

We summarize here the equations for the boundaries of the Arnol'd tongues in
the two $a$-intervals. We denote these respectively by $i=0$ and $i=1$: 
\vspace{-2ex}
 \bna \quad i=0: 0 \leq a \leq 1   \mbox{\qquad and \qquad} 
i=1: 1 \leq a . \ena
\vspace{-4ex}

The equations for these boundaries  
in the non-chaotic interval $0 \leq a \leq 1$  (the case $i=0$) 
have been determined by  Ding and Hemmer
\cite{Ding}. They are, for easy access,    reproduced here also.  

 Those in the second, chaotic $a$-interval 
have been obtained by one of the authors~\cite{JG:tbp}. 

 We   now first 
introduce some further notations:

For a given, fixed rational number $J=q:p$ and $i=0$ or $1$, we write the
equations for these boundaries in the form

\be b_i^-(a) \leq b \leq b_i^+(a) \Label{bds:b}, \ee

whereby it is understood that they hold true for $b$ -values such that the
difference $\Delta b_i(a)$ defined by

\be \Delta b_i(a) = b_i^+(a) - b_i^-(a) \ee

is nonnegative.  In addition to these quantities $\Delta b_i(a)$ we also
introduce the following notation for the midpoint $\bar{b}_i(a)$ of such a $b$
interval. i.e.\, we define this quantity by

\be \bar{b}_i = \half( b_i^+ + b_i^-). \ee

It is clear that, in order to know the bounds $b_i^\eps(a)$ it is sufficient
to know $\Delta b_i(a)$ and $\bar{b}_i(a)$.

\subsection{The Ding-Hemmer formula for the case $0 \leq a
\leq 1$ } \Label{ssect:DHrslt}

The result of Ding and Hemmer \cite{Ding} can now be formulated as follows:

0.1) Their result implies for $\bar{b}_0(a)$:
 \be 2\bar{b}_0 (1-a^p) / (1-a) = 2q - 1 + a^{p-1} - 2 (1-a) \Dc /a
\Label{eq:BrB-DH} \ee
 
where $\Dc$ is the polynomial in $a$ defined by

\be \Dc \equiv \Dc(a) = \sum_{n=1}^{p-1} [\frac{nq}{p}] a^{p-n}
 \ee

and $[x]$ is the `entier function' defined by

\be [x] \in \ZZ, \quad 0 \leq x-[x] < 1 .  \ee

0.2) Their result implies for $\bar{b}_0(a)$:

\be \Delta b_0(a) = \frac{(a-1)^2 a^p}{a(1 - a^p)}\Label{eq:DltBt0}. \ee

\subsection{The result for the case $a \geq 1 $} \Label{ssect:NRslt}

Also for the case $i=1, a \geq 1$ the equations for the boundaries of the
Arnol'd tongues have now been obtained \cite{JG:tbp}.  They can be summarized
in the form:
 
1.0)

\be \bar{b}_1(a) = \bar{b}_0(a) \Label{eq:barb} \Label{eq:barb2} \ee

1.1)

\be \Delta b_1(a) = \frac{(a-1)^2 (2 -a^p)}{a(a^p-1)} . \Label{eq:Delta-1b}\ee

We note that, although the boundary curves of the two Arnol'd tongues with the
same value of the current but lying on different sides of the line $a=1$ are
given by different equations, the equations for the quantities $\bar{b}_i(a)$
and $\Delta b_i(a)$ in terms of which these boundary equations can be
expressed, are closely related.  The first of these relations is the above
Eq.\ (\Ref{eq:barb2}), and the second is  the following proportionality
relation:

\be \frac{\Delta b_1(a)}{(a^p-2)} = \frac{\Delta b_0(a)}{a^p} .
\Label{rel:ratio} \Label{rel:prop} \ee
 
\subsection{Corollaries of these formulas} \Label{CorrlrsA}

 From Eq.\ (\Ref{eq:Delta-1b}) one finds, as can also be derived by a simple
argument, that a tongue, with $a\geq 1$, characterized by integers $p$ 
and $q$ according to
the above convention, will have an   intersection of  positive 
length  with a line of
constant $a$ if and only if $a$ satisfies

\be 1 < a^p < 2 \Label{bds:ap} \ee

implying that, in a plot of $D(a,b)$ at constant $a$ with $a>1$, the
collection of finite-length intervals with vanishing $D$ which occur are
exactly those which have a current $J$ equal to a rational value $J=q:p$ with
$p$ satisfying the above inequality (\Ref{bds:ap}).

And there will be no other intervals with vanishing $D$ because, outside any
one of the Arnol'd tongues, $D(a,b)$ is a fractal function of $b$ which cannot
vanish -- identically in $b$ that is -- in any finite-length $b$-interval.

\section{Response: macroscopic and microscopic }\Label{sect:Resp}
   
Here we list the various types of response we have found reason to 
distinguish  in our model:

1. Negative Macroscopic Response

2. Fractal Nonlinear Response

\subsection{Macroscopic response} \Label{ssect:RespMa}

As mentioned, a striking feature of our model is that, when the parameters $a$
and $b$ are chosen in the right range,  there is a good chance that for the
response to be  negative.

There are two versions of this effect:

I) The current $J$ and the bias $b$ may have opposite signs, i.e.\
$J(a,b)/b<0$. This is felt as counter-intuitive and would need an explanation.
 
It occurs most frequently when $b$ is rather small, and $a$ just above any odd
integer, see Figs.\ \Ref{E5} and \Ref{E7}.

II) The `dual' version of the same effect is when $\half - J$ and $\half - b$
have opposite signs.

It is just as `counter-intuitive' as the first version of the effect, as can
be seen most clearly by performing a coordinate transformation on the $x$-axis
$x \to \half - x$ and similarly replacing $b$ by $\half - b$.
 
That two such versions of the effect occur may be seen to be correlated with
the fact that a `fundamental interval' for the $b$ parameter of the model has
two symmetry points, which are at the end points of the `fundamental interval'
$0 \leq b \leq \half$.

For the same reason the `fundamental strip' chosen for the $(a,b)$ plane is a
closed region bounded by the two straight lines given by these $b$-values.
 
To study these two effects more closely we have plotted both effects,
respectively, in Fig.\ \Ref{E5} and Fig.\ \Ref{E6}, and as it has come out,
these plots look quite similar to each other apart from an interesting shift
in the $a$ direction over a unit distance.

These two effects become more and more pronounced the closer one gets to the
respective symmetry line and for a moment could even be thought of as being in
conflict with some basic law of Statistical Mechanics.  However, the latter
can of course not be true because the dynamics is not
Hamiltonian. Nevertheless, this feeling of counter-intuitiveness remains,
calling for a better explanation of the effect or a better understanding of
what precisely that intuition would tell us.

 From the Figures \Ref{E5} and \Ref{E6} it looks as if, in each of two these
`versions' of the effect, the boundary has a nontrivial 
structure on any sufficiently small scale
and hence \cite{Beck} should be considered a `fractal set'.

 We note that phenomena somewhat   similar to our `negative currents' is
observed in certain model systems known as `ratchets', where typically the
word `current reversal' is used (Cf., e.g.\, \cite{Reimann}).
 
\subsection{Microscopic nonlinear response} \Label{ssect:NLR}

The strong, even fractal, nonlinearity of the various responses found in our
model brings to mind, after a long period, discussions taking place concerning
the range of validity of the hypothesis of Linear Response.

Let us recall that Linear Response  as such is an experience of everyday
life, which since day and age has found its expression in countless
phenomenological laws of physics, such as Ohm's law, Fick's law and many other
ones.

A derivation of such linearities, under quite general circumstances, directly
from the laws of Statistical Mechanics of many-particle systems, was put
forward by Kubo \cite{Kubo}, whose theory, or `hypothesis' as we like to call
it, has since then become a fundamental and by now well-established
\cite{West} part of the Nonequilibrium Statistical Mechanics of many-particle
systems.
  
However, already at an early stage, Van Kampen \cite{vKI} expressed his 
concern about the lack of mathematical rigour of the derivation and has put in
doubt the general validity of the hypothesis.

That Linear Response cannot be valid in complete generality has since then
been proven by the discovery of certain counter-examples, one of which is that
of the non-existence of the usual hydrodynamical equations in two
dimensions.\footnote{We owe this remark to H. van Beijeren.}

Therefore, general conditions, sufficient or necessary, under which the
hypothesis of Linear Response is valid, seem to be to this moment not
known and it is here, when an attempt should be made to clarify this issue,
that our present model might prove useful.

In fact, the model provides a clear scenario as to how Linear 
Response  can be violated, a
scenario which might also be present in more realistic cases. It can even not
be excluded that strong effects similar to the ones found in our present model
could occur also in Hamiltonian systems.

\section{The figures} \Label{sect:Figs}

In this section we will discuss in more detail the figures of this paper.\\
{\bf Figure \Ref{E1}:} 
Projected three-dimensional plot of (a) the current $J(a,b)$ and (b) the
diffusion coefficient $D(a,b) $ as functions of the system parameters $a$ and
$b$. One may notice, in part (a), close to the upper corner of the graph at
$a$ just above $2$ and $b$ just below $\half$, the `bump' or little `hill'.
The existence of such a local maximum of $J$ implies, as is not difficult to
see, that what we have termed our `alternative ratio', $(\half - J)/(\half -
b)$, is negative in that region.
 
As we already explained, the relation between these two effects may lose some
of its mystery by realizing that both $b=0$ and $b=\half$ are `symmetry
points' of the $b$-symmetry group of Subsection \Ref{ssect:Sym}, in the sense
that each is a fixed point of a reflection subgroup of that symmetry
group. This analogy between the two effects can also be seen simply by
changing the $x$-coordinate and the $b$-parameter simultaneously over a
distance $\half$.\\
{\bf Figure \Ref{E2}:}  
Graphs of the diffusion coefficient $D$ as a function of $b$ at constant $a$.
In parts (a)-(c) the constant values of $a$ are respectively $a=1.125$,
$a=1.075$ and $a=1.0375$.
 
These values are chosen so as to approach, in an approximately geometric
fashion, the transition line at $a=1$ between the non-chaotic and one of the
two chaotic regions (the other  chaotic region being   located at $a<-1$).

In part (d) these three graphs are superimposed, in a close-up containing in
each case the interval of vanishing $D$ around $b=1/3$.

On these graphs one may observe: 

(1.)  As $a$ decreases towards the limiting value $1$ where the transition to
non-chaotic behaviour takes place, the number of intervals with $D=0$
increases rapidly whereas separate plots of $J$ (not shown here), show that
inside such intervals $J$ has a constant, always rational value.
 
(2.)  Outside of these intervals, $D(a,b)$ shows fractal behaviour as a
function of $b$.

(3.)  Around these intervals, $D(a,b)$ shows spikes, which usually is an
indication of a critical behavior of some sort. However, from the continuity
of $D(a,b)$ throughout the chaotic region $|a|>1$ \cite{JG:tbp} it follows
that $D$ cannot become infinite and could 
only become unbounded when approaching
the  line $a=1$. This  is not the case in this figure, so that the
observed `spikes' must be finite. This, however, leaves open the possibility
that some derivative of $D$, in as far as it may exist, would become infinite.

Elaborating a little further on the first point (1.) above, one may verify in
more detail that each of these observed intervals of vanishing $D$ is indeed
determined by the intersection of the line of constant $a$ with a respective
Arnol'd tongue. This can be verified as follows:

The formulas for the Arnol'd tongue boundaries presented in Section
\Ref{sect:Arnld} imply that such a tongue with $J=q/p$ in the standard
notation will extend between $a=1$ and a maximum value of $a$ given by
$a=2^{1/p}$ so that a line of constant $a$ will intersect those and only those
Arnol'd tongues which have a $p$-value satisfying $a^p < 2$.  For the three
graphs shown in the parts (a)-(c) of this figure this means that the maximum
values of $p$ are respectively $5$, $9$ and $19$.  With this information it is
then easy to identify, in these three graphs, for each interval of vanishing
$D$ the rational value of $J=q:p$. For example, in part (a) one will in this
way identify for the successive observed intervals with $D=0$ the following
values of $J$ respectively: $J=0:1, 1:5, 1:4, 1:3, 2:5$ and $1:2$.  A similar
identification is possible for the other two graphs.

In other words, the graphs of this Fig. 2 nicely illustrate the theoretical
explanation given in Section \Ref{sect:Arnld} for the existence of Arnol'd
tongues in this model.\\
{\bf Figure \Ref{E3}:}
Graphs of the current $J(a,b)$ as a function of $a$, at constant $b$, for
three different values of $b$.  In part (a): $b=0.1$, in part (b): $b=0.01$
and in part (c): $b=0.001$.

Again, as in the case of $D$ in the preceding figure, a highly irregular
behaviour of this function emerges, which becomes wilder and wilder the closer
$b$ gets to $0$.

One may notice here, on comparing these three graphs, that, only roughly
speaking since $J$ has so much variation in it, each time $b$ is scaled down
by a factor $10$, $J$ also scales down but by a smaller factor; as was already
necessary in order to keep all points of the curve inside the picture.
This by itself is already indicative of nonlinear behaviour in $b$, which,
however, is not so simple to describe by a single exponent since so much
appears to depend on the precise value of $a$ at which one lets $b$ approach
to $0$.
 
This $b$-dependence of $J(a,b)$ as $b$ approaches to $0$ appears to depend
quite sensitively on the value of $a$.  Some of this can be made more precise.
For example, as can be proved from e.g.\ Eq.\ (\Ref{expr:c1g}), for $a$ equal
to an odd integer, $J=b$ exactly.

In the general case one can say something about the range of limiting values
taken by $J(a,b)/b$ as $b \to 0$: For general $a$ it can be shown to be
unbounded, whereas the ratio $J(a,b)/(b (|\log|b|))$ remains bounded but
has, for a general choice of $a$, no limit.  More details will be given
elsewhere \cite{JG:tbp}.
 
We mention one other observation which can be made on comparing these three
graphs. It is that, as $b$ decreases towards $0$, there appear more and more
intervals on which $J/b$ is negative.\\
{\bf Figure \Ref{E4}:}
Here, the ratio $J(a,b)/b$ is displayed as a function of ${}^{10}\log b$, for
$b$ ranging through eight decades from $b=1$ on downwards.  This is at the
three different constant values of $a$ as indicated in the figure.

For each of these three values of $a$ the current $J$ shows a highly irregular
behavior as a function of $b$, persisting, upon enlargements of the graph, on
finer and finer scales, which is indicative of a fractal structure of $J(a,b)$
as a function of $b$.
 
Also, one observes in all three graphs that the ratio $J/b$ is negative over
rather large $b$-intervals.  To our knowledge, this is the first finding of
`negative currents' in simple piecewise linear one-dimensional maps.
Therefore, both effects, that of negativity of $J/b$ and nonlinearity of $J$
versus $b$, show up in these graphs, and there is no indication that
eventually, from a certain small value of $b$ on, this behaviour will
disappear and a regime of linear response will be reached, i.e.\ where $J/b$
would approach a constant.
  
But the situation is a bit more complicated: There are special values of $a$,
e.g.\ if $a$ is an integer, that then, quite trivially, $J(a,b)=b$ identically
for all $b$.
    
Also, this should not be confused with the `large field linearity' of $J(a,b)$
meaning that $J$ will become asymptotically equal to $b$, a property which
follows in an elementary way from the fact that $J(a,b) - b$ is periodic and
hence bounded so that $J(a,b)/b$ will tend to $1$.  This holds quite generally
under quite weak conditions for lifted circle map systems. But here at least
it can be concluded that, as $b$ increases, $J(a,b)$ will eventually become
positive and remain so.

We also note that the three values of $a$  were  chosen here so as to let $a$
approach, in a roughly geometric fashion, to 
the integer value $a=3$, which is a
value for which a plot of the ratio $J/b$ would show no irregular behaviour at
all, since this ratio then is identically equal to $1$. The observation
therefore is that, the closer one gets to a regular point such as $a=3$, the
more pronounced the singular behaviour seems to become; but that is has
disappeared completely when the limiting point has been reached.\\
{\bf Figure \Ref{E5}:} 
Various close-ups of parts of the parameter plane where the phenomenon of
`macroscopic negative response' i.e.\ where $J(a,b)/b<0$ takes place.  In part
(a), the regions where $J(a,b)/b<0$ are depicted in grey, and their boundaries
in black. In parts (b) and (c) further enlargements of these regions are given
where only the boundaries with $J(a,b)$ changing sign are depicted.  One
notices the self-similar structures which become visible; whose  
precise nature
 however is not  clear yet.\\
{\bf Figure \Ref{E6}:} 
This figure is similar to Fig. {5} except that here the sign of the
`alternative' ratio $(\half - J(a,b))/(\half - b)$ is mapped out, instead of
that of $J/b$. In these graphs, only the boundaries of the regions
of constant sign of this `alternative' ratio, i.e.\ the curves where
$J(a,b)=\half$, are depicted.\\
{\bf Figure \Ref{E7}:}
 
This figure needs a longer than usual explanation because of the many
details of so many different kinds it contains.  We term it our
`chart', as it displays broadly speaking the `qualitative features' of
the system we have found in our preliminary survey.  These features
are all integer-valued which makes this two-dimensional chart
possible. They are of four different types, labelled from I to
IV. Those labelled I, II or III are clearly invariants of a
respective, well-known invariance group, which will be indicated
below.  Also in the case IV a similar association seems possible.
 
 Because of the $b$-symmetries discussed in Subsection
\Ref{ssect:Sym}, we have, without lack of generality, restricted our
attention to the strip $0 \leq b \leq \half$, termed there the
`fundamental' strip of the parameter plane.

The information displayed in this figure is in the form of

a) the brackets containing three integers;
 
b) various kinds of lines and curves, and 

c) the shaded areas near the two boundary lines of the graph, at $b=0$ and
$b=\half$.

As for a): the pair of integers on top are the local values of the kneading
numbers $n_1^+$ and $n_1^-$ whereas the integer on the bottom is the number of
fixed points $\mbox{Fix}(f)$ of the associated circle map $f$ (Cf. Section
\Ref{sect:CGeom}).

As for b): Each curve or line is  where a respective
kneading number changes its  value.

\qquad b1) The sequence of dashed straight lines are in that way `indicators'
of one particular sequence of the first order kneading numbers. 
In the  sequence of straight lines, when it has entered  
the non-chaotic
region $0 \leq a \leq 1$, the last one has become 
a boundary of an Arnol'd tongue, the one with $J=0, p=1,q=0$.

\qquad b2) The `zig-zagging' sequence of curves, nearly parallel to
the straight lines of b1), are `indicators' in the above sense of a similar
sequence kneading numbers, this time of order 2.  As this sequence has  
entered 
the non-chaotic region, the last one has  become 
a boundary of the Arnol'd tongue characterized by $J= 1:2, p=2, q= 1$.
 
\qquad b3) Most of the other curves indicate boundaries of Arnol'd
tongues.  Those which are displayed prominently,  in the chaotic as well 
as in the non-chaotic regions, are the complete sets with $p$-values 
ranging from $p=1$
up to $p=5$.  These have been plotted using the exact equations for these
boundaries of  Section \Ref{sect:Arnld}.

\qquad b4) Near the line $a=1$ also the boundaries of in principle all
higher order (and much smaller) Arnol'd tongues are displayed. These have been
plotted using the computer program implementing formula Eq.\ (\Ref{expr:c1g}).

\qquad b5) The dotted vertical line at $a=1$ denotes the boundary
between the chaotic and the non-chaotic regions, which is where the sign of
the Lyapunov exponent, an invariant of Type III, changes.
 
As for c): The shaded areas near the boundaries at $b=0$ and $b=\half$ are
where either one of the two `macroscopic' responses is  negative. These
regions are also shown, on much larger scales, in Figs.\ \Ref{E5} and
\Ref{E6}.

In the list below we summarize in which way, i.e.\ by
which ones of the above
listed signs (a) -- (c), the information on the values of some of the
invariants of the four types I-IV is displayed:

(I, $p$ and $q$ of Section \Ref{sect:Arnld} ; Ergodic Theory) : b3, b4.

(II, Kneading numbers, Order Topology) : a, b1, b2, b3.
 
(III, Sign of Lyapunov exponent): b5.

(IV, The  signs of the two current-to-bias ratios): c.

\section{Further problems and outlook}\Label{sect:Disc}

One of the features of our model of which we would like to obtain a better
understanding, from a `physical' or `probabilistic' point of view, would be
the mechanism which is responsible for the ` negative response' observed in
certain regions of the parameter plane. One approach would be to make
a further, mathematical analysis of the explicit but subtle formula
Eq.\ (\Ref{expr:c1g}) for $J$ of Section (\Ref{sect:Sol}) determining
that sign.

Another approach would be to take advantage of the connection
\cite{RK:tbp} between the present model and certain types of ratchet
models \cite{JuKiHa} in which `negative currents', or, in the
terminology used, `current reversals', also occur. The dynamical
origin of the negative currents in these models is currently under
discussion \cite{Mateos,BarbiSalerno}. For a recent review on ratchet
models see, e.g.\ Ref.\ \cite{Reimann}.

\section{Summary} \Label{sect:Summ}

We have investigated a simple two-parameter model of chaotic dynamical
transport, along lines of earlier investigations
\cite{RKD,RKdiss,KD99}, but this time using as our principal tool the
exact expressions for the transport properties $J$ and $D$ obtained
recently \cite{JG:tbp}.

These formulas are explicit and allow for a highly efficient (`polynomial
time') computation, but analytically they are quite subtle, and the functions
they represent have a fractal character. For this reason it was necessary to
implement them numerically, in order to obtain at least a reasonable
impression of the various properties of the model.
  
 In the course of this both numerical and analytical investigation
some unexpected features of the model then came to light, as are
displayed here in several figures and are amply discussed.
 
Our most significant findings are of two kinds:

1. The ubiquitous continuous but fractal parameter dependence of every
one of the `near-equilibrium' transport coefficients, such as the
current $J=c_1$ and the diffusion coefficient $D=c_2$, throughout the
chaotic part of the parameter plane, naturally with the exception of
the Arnol'd tongues where they are constant anyway.  Only one
particular aspect of this is the `fractal nonlinearity' of the current
$J(a,b)$ as a function of $b$, implying that, for most $a$-values, the
limit of $J/b$ as $b\to 0$ does not exist.

2. A second, hitherto unexpected and thus far counter-intuitive
feature of our model is the negativity of the ratio $J/b$ in many
regions of the parameter plane.  This effect has two complementary
versions, each relative to its respective symmetry line in the $(a,b)$
plane.  These effects occur in irregularly shaped regions which are of
positive measure; and which regions close in onto the respective
symmetry line, thereby showing critical behaviour.\\
 {\bf Acknowledgments}
The authors are indebted to J.R.  Dorfman in many ways:
 
It has each time been a privilege for J.G.\ to attend, during the years, the
lucid and inspiring lectures by Professor Dorfman on the latest and exciting
developments in Non-Equilibrium Statistical Mechanics.  On one of these
occasions, May 1994, he introduced a beautiful simple model, of which the one
discussed in the present paper is an extension, exhibiting fascinating fractal
curves, and which shortly thereafter turned out to be exactly solvable.  The
author wants to thank him for these lectures, for his kind interest in the
present work, and for his warm hospitality during a visit, December 1996, to
the University of Maryland where that solution was discussed in detail for the
first time and presented in a seminar.  J.G.\ also wants to thank his
colleagues and former colleagues of the Institute for Theoretical Physics in
Utrecht, in particular N.G. van Kampen and Th.W. Ruijgrok, for many valuable
discussions on problems of physics and mathematics.

R.K. thanks J.R. Dorfman for being a wonderful teacher of physics during many
years of exciting joint research on deterministic chaos and transport.
Indeed, this collaboration started with R.K.'s Ph.D.\ thesis work on the
symmetric case of the model discussed above, as initiated and supervised by
Prof. Dorfman.  This author furthermore wants to thank especially H.  van
Beijeren for his strong support of this work, as well as M.H.  Ernst, P.
Gaspard, G.  Nicolis, G.  Radons, and H.  Spohn, for further support and
encouragement.  R.K.  is currently a PKS fellow at the Max Planck Institute
for Physics of Complex Systems.

\renewcommand{\baselinestretch}{1}

\begin{figure}[t]
\epsfxsize=10cm
\centerline{\epsfbox{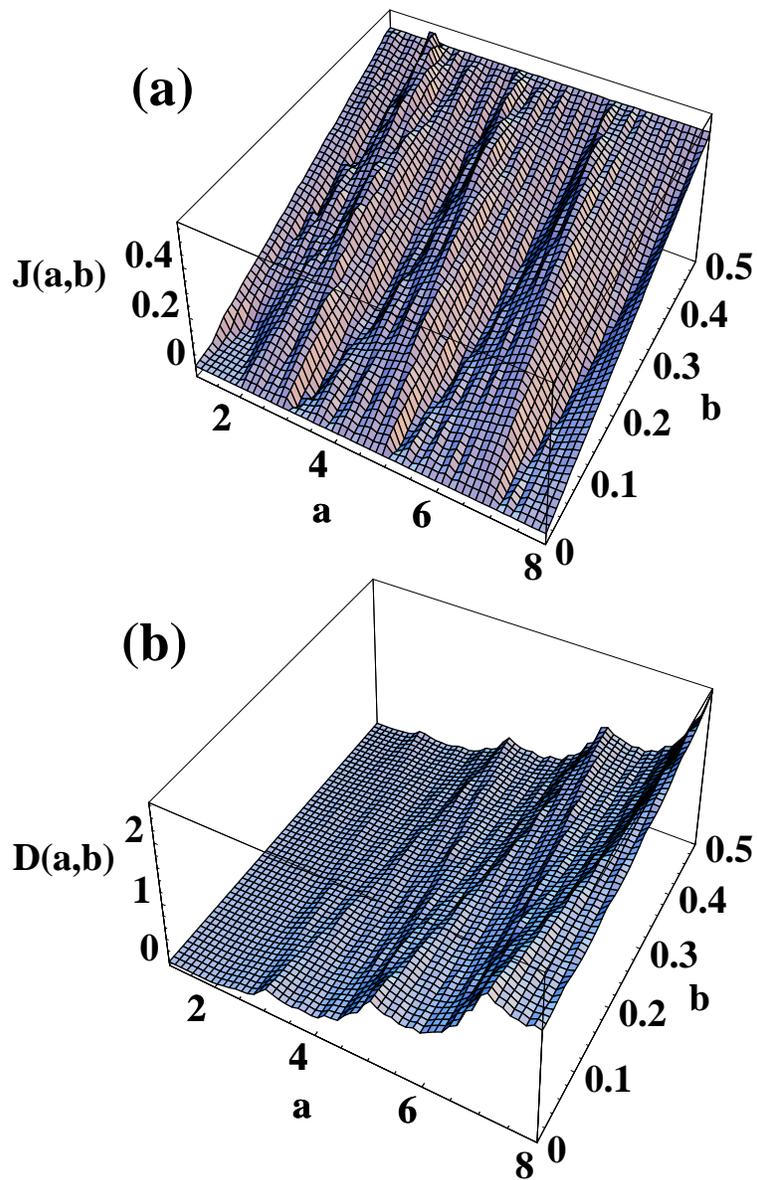}}
\vspace*{-2cm}
\caption{\Label{E1} Projected three-dimensionsal plots of 
the current $J(a,b)$ and the diffusion coefficient $D(a,b)$, as functions of
$a$ and $b$.}
\end{figure}

\begin{figure}
\epsfxsize=13.5cm
\centerline{\rotate[r]{\epsfbox{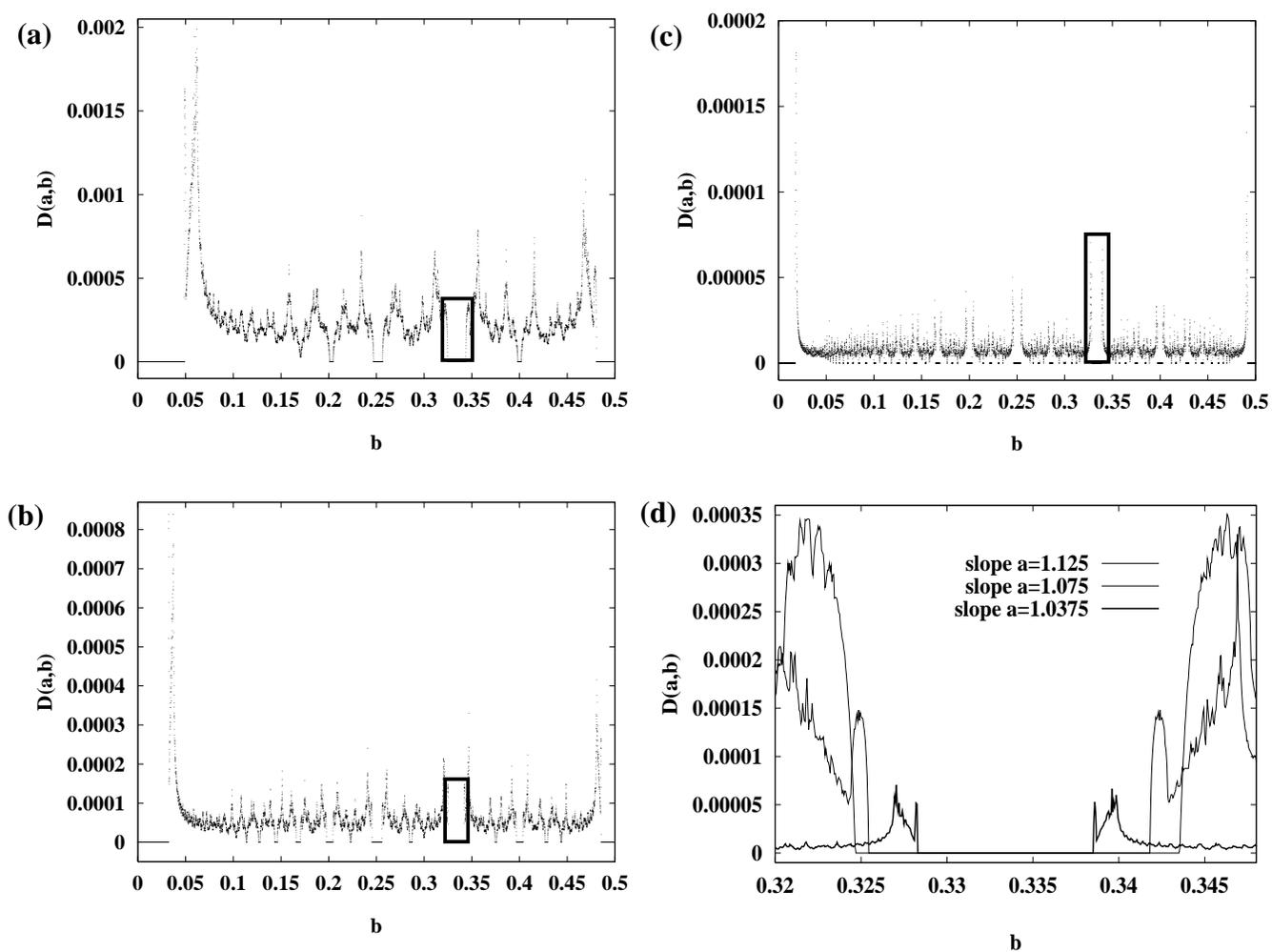}}}
\caption{\Label{E2} Graphs of the diffusion coefficient $D$ as a function 
of $b$ at constant $a$.  In parts (a)-(c) $a$ is held constant at respectively
the values $a=1.125$, $a=1.075$ and $a=1.0375$.}
\end{figure}

\begin{figure}
\epsfxsize=10cm
\centerline{\epsfbox{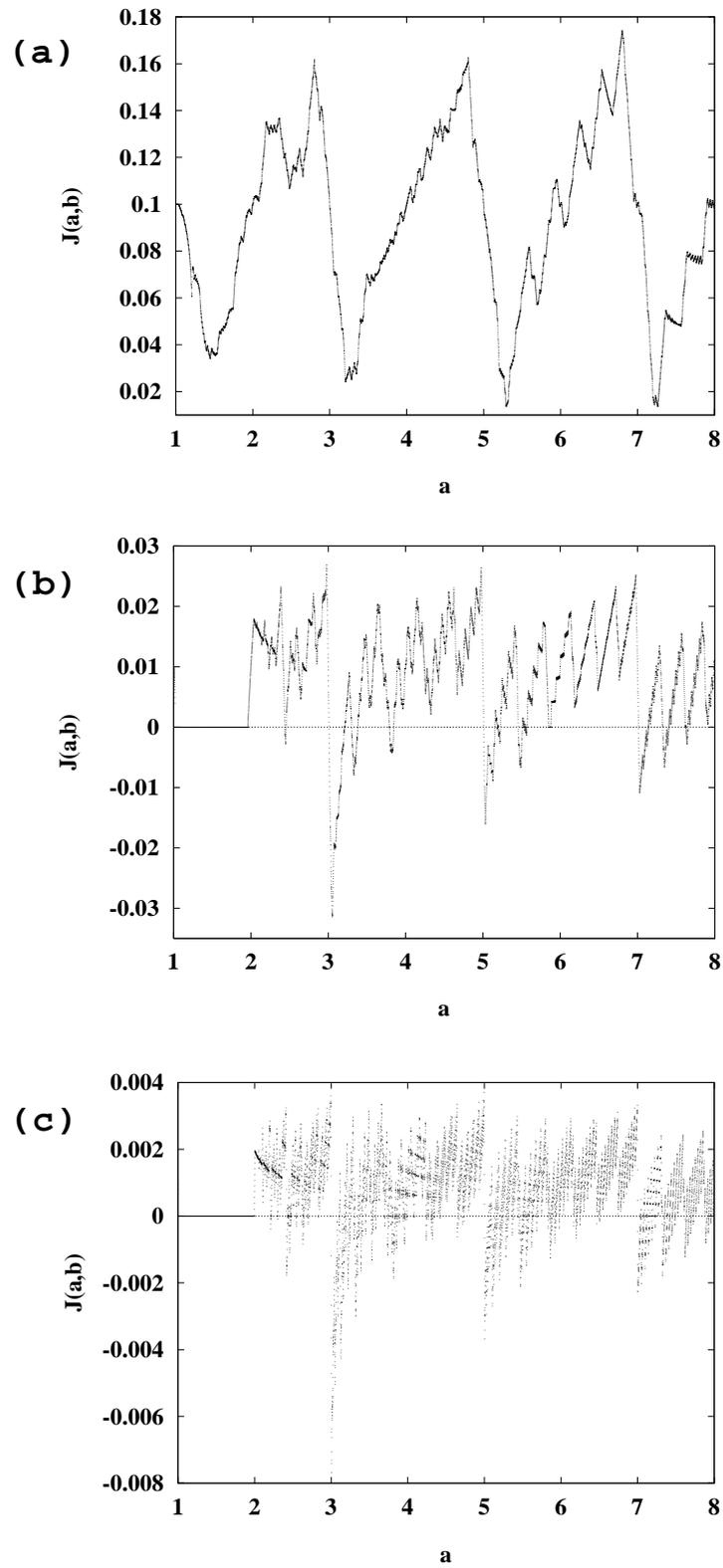}}
\vspace*{0.7cm}
\caption{\Label{E3} Graphs of the current $J(a,b)$ as a function of $a$ at 
constant $b$. In parts (a)-(c) $b$ is held constant at respectively the values
$b=0.1$, $b=0.01$ $b=0.001$.}
\end{figure}

\begin{figure}
\epsfxsize=13cm
\centerline{\rotate[r]{\epsfbox{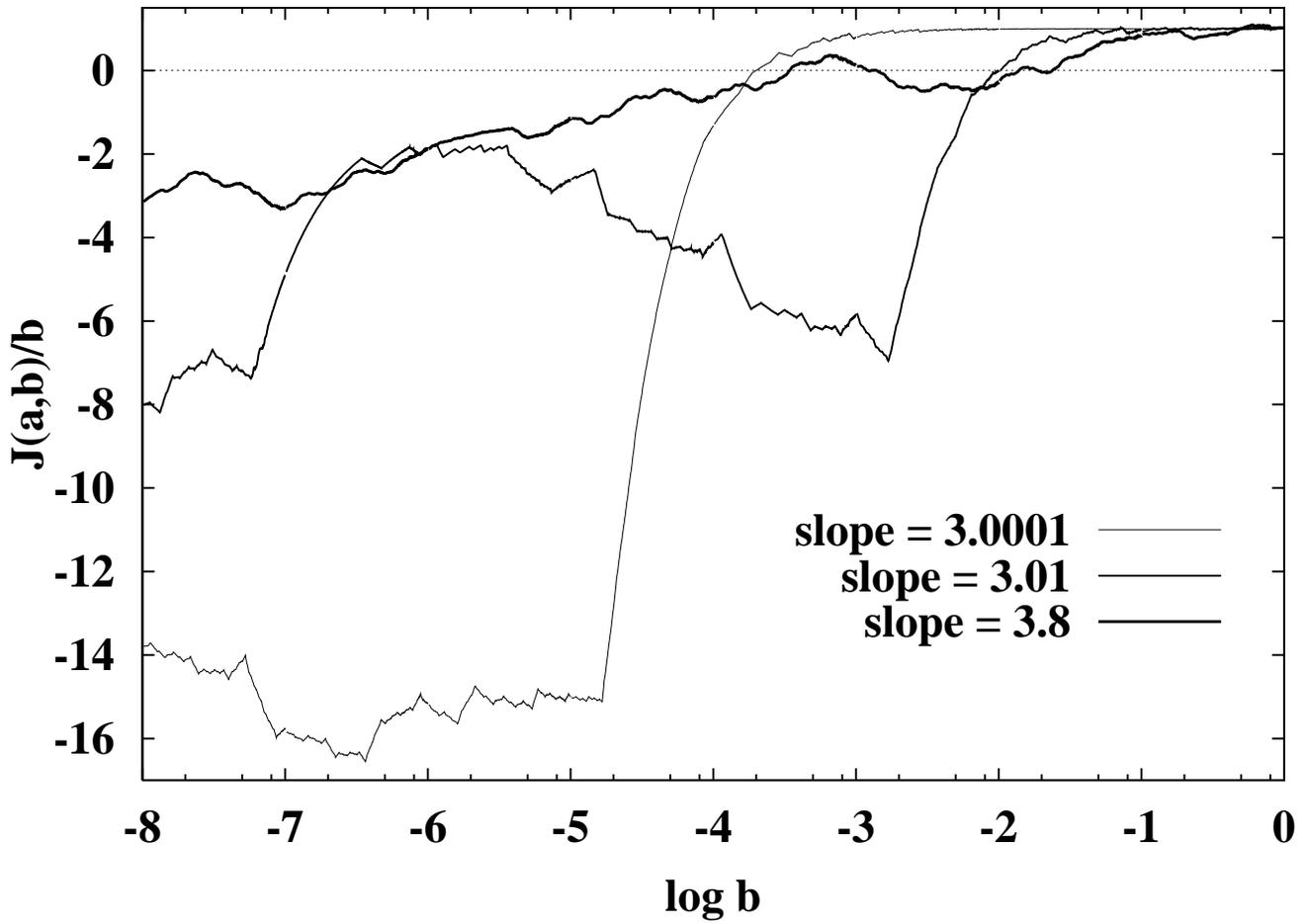}}}  
\vspace*{0.5cm}
\caption{\Label{E4} The ratio $J(a,b)/b$ as a function 
of ${}^{10}\log b$, with $b$ ranging over eight decades.  Herein, $a$ is held
constant at the three different values as indicated in the figure.}
\end{figure}

\begin{figure}
\epsfysize=20cm
\centerline{\rotate[r]{\epsfbox{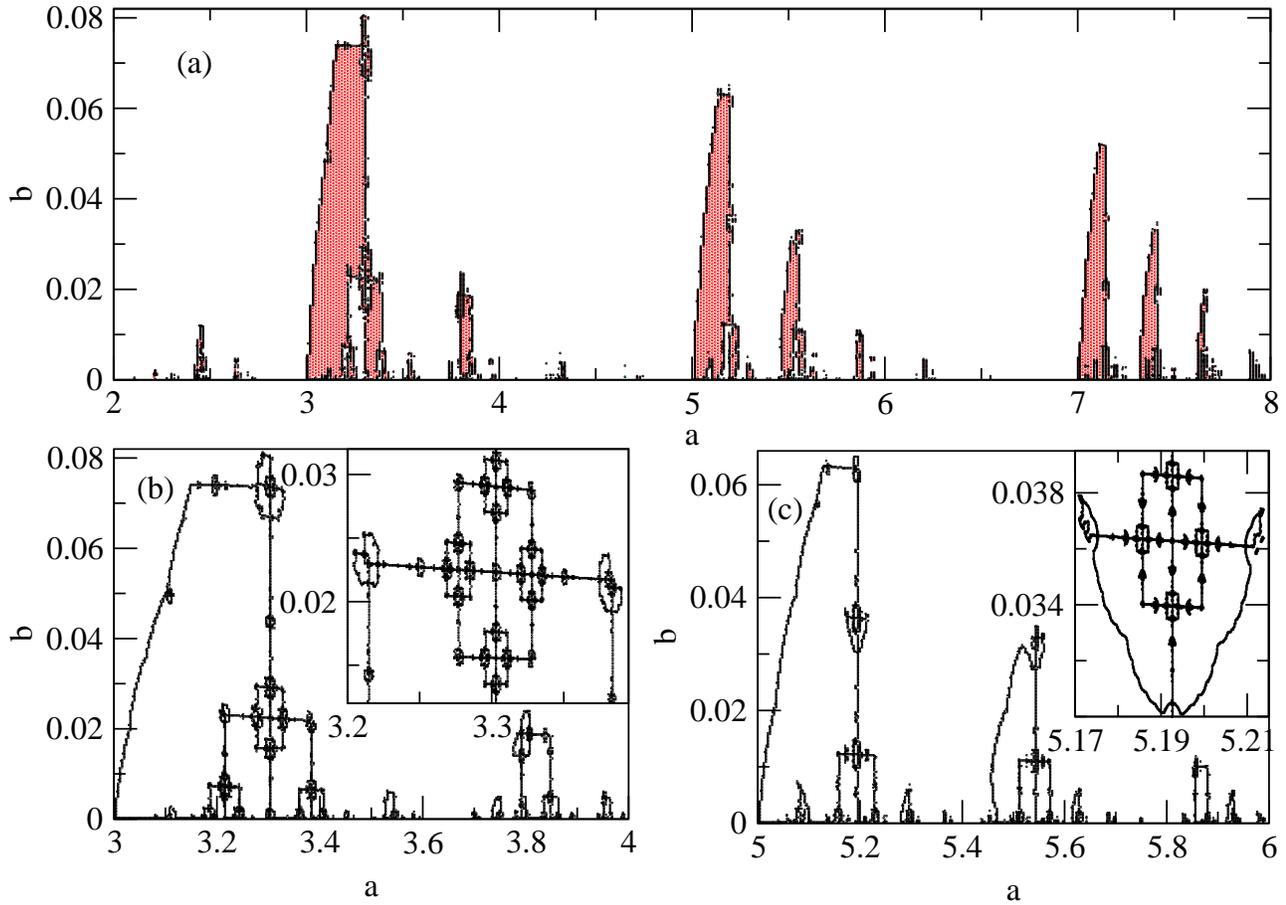}}}
\caption{\Label{E5} Various close-ups of parts of the parameter plane where
`macroscopic negative response' occurs, i.e.\ where $J(a,b)/b<0$.  In part
(a), regions with $J(a,b)/b<0$ are coloured grey, and the boundaries thereof,
which is where $J$ changes sign, are in black. Parts (b) and (c) are further
enlargements displaying now only these boundaries. Each contains an inset with
a further enlargement, making the self-similar structures present in these
boundary curves visible.  Their precise nature is not yet clear.}
\end{figure}

\begin{figure}
\epsfxsize=15cm
\centerline{\rotate[r]{\epsfbox{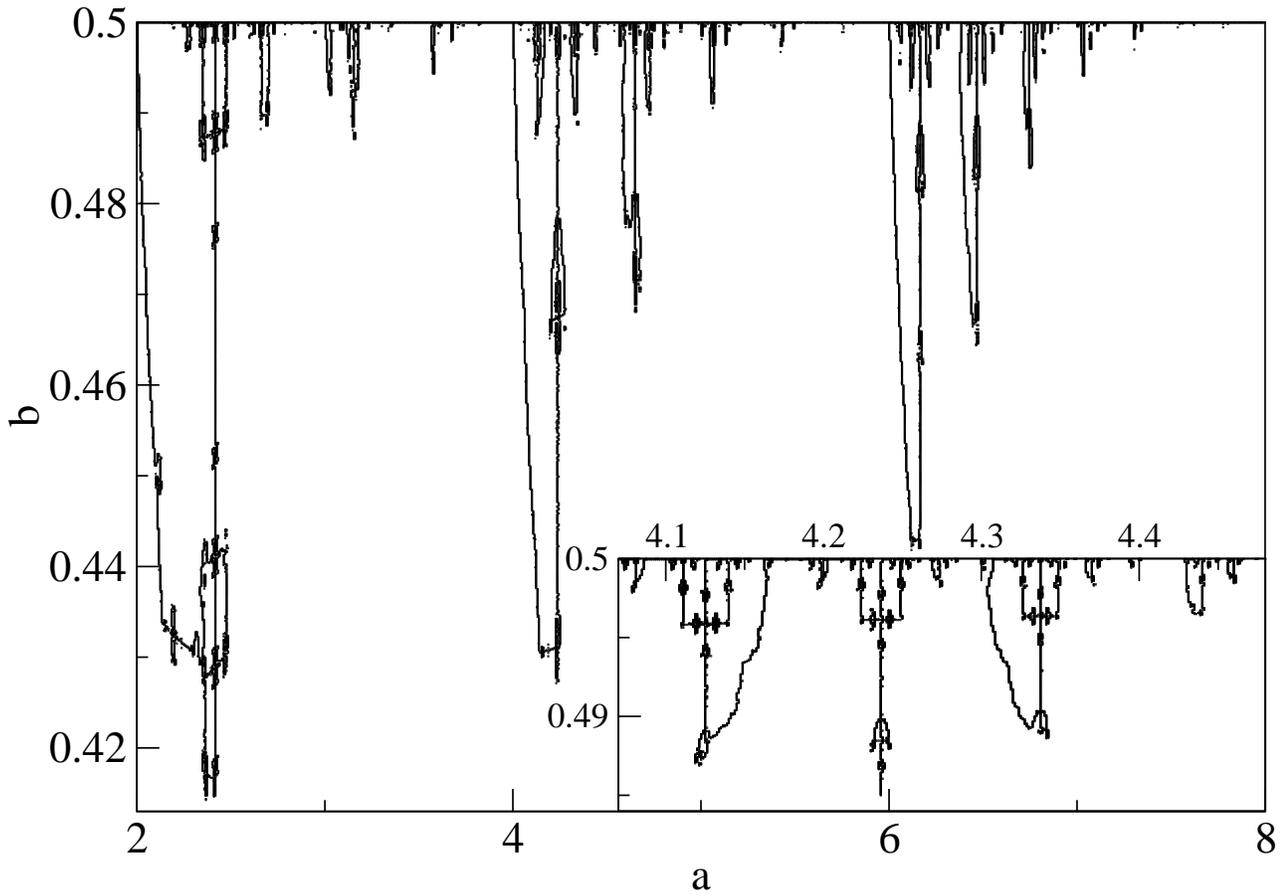}}}
\vspace*{0.5cm}
 \caption{\Label{E6} This figure is  quite similar to Fig. \Ref{E5} except 
for two differences:
 1) These curves display the `alternative' quantity $\half -
J(a,b)$ rather than the quantity $J(a,b)$ as in the previous figure.
 2) The $b$-scale used is different.
 Taking the latter into account one observes that this negativity
effect starts off here earlier, i.e. around $a=2$ rather than at
around $a=3$, and is then accordingly also larger, as compared to the
case of Figure \Ref{E5}.}
  \end{figure}

\begin{figure}
\epsfxsize=15cm
\centerline{\rotate[r]{\epsfbox{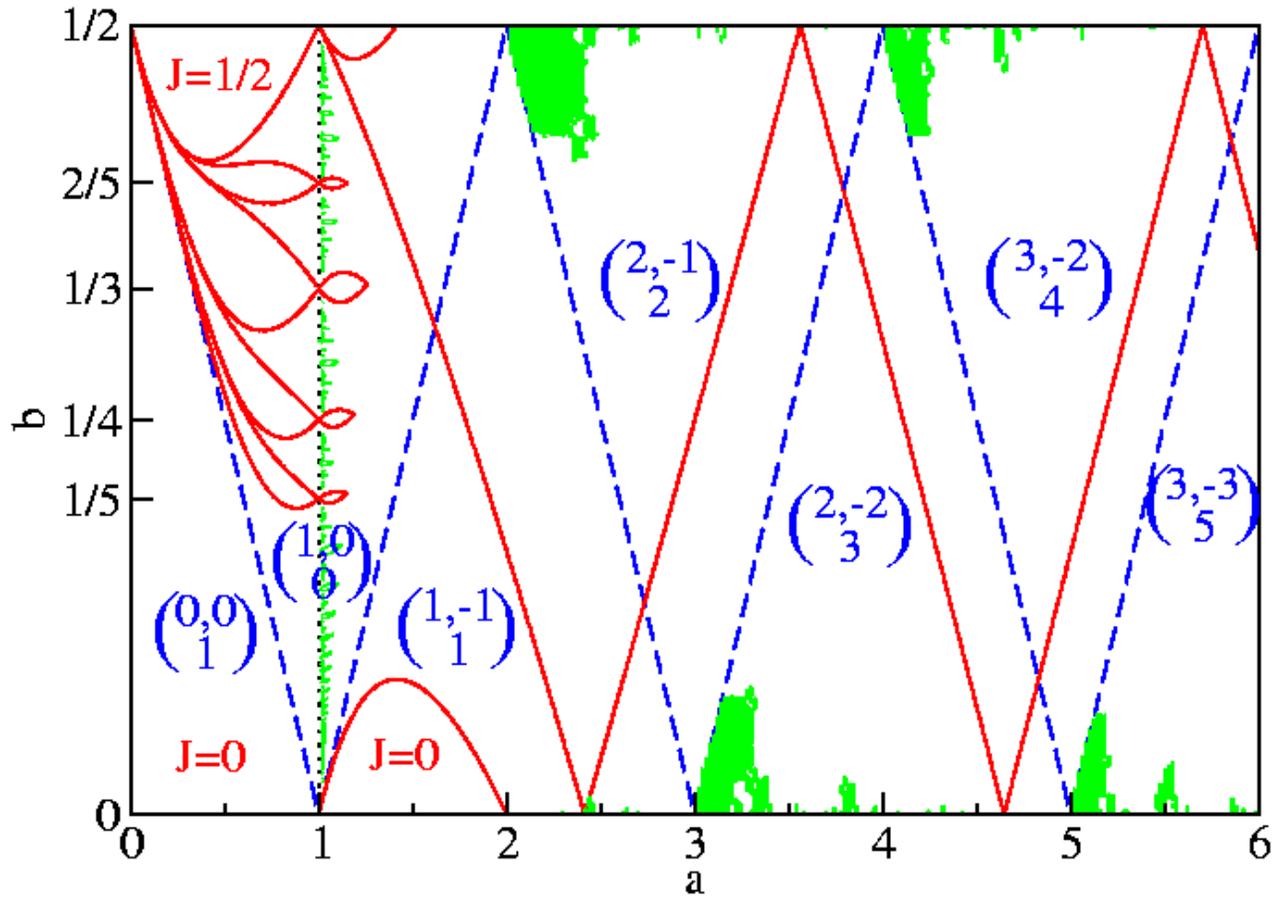}}}
\caption{\Label{E7} This figure, which is our `chart' of the model, displays 
values of invariants which are of four different types. For a detailed
explanation of the figure itself see Section \Ref{sect:Figs}.}
\end{figure}

\end{document}